\documentclass[12pt,tightenlines,eqsecnum,floats,showpacs,nofootinbib,amsmath,amssymb,aps,prd]{revtex4-2}

\usepackage{tikz}
\usepackage{graphicx,verbatim}
\usepackage{amsmath}
\usepackage{amsfonts}
\usepackage{amssymb}
\usepackage{psfrag}
\usepackage{accents}
\usepackage{mathrsfs}
\usepackage{enumitem}
\usepackage{hyperref}

\newcommand*{\scri}{\ensuremath{\mathscr{I}}} 
\newcommand*{\scrip}{\ensuremath{\mathscr{I}^{+}}} 
\newcommand*{\scrim}{\ensuremath{\mathscr{I}^{-}}} 

\usepackage[colorinlistoftodos]{todonotes}
\usepackage{epstopdf}

\def\be{\begin{equation}}
\def\ee{\end{equation}}
\def\ba{\begin{eqnarray}}
\def\ea{\end{eqnarray}}

\newcommand{\pb}[1]{\hbox{\lower0.5ex\hbox{${}_{\leftarrow}$}}\kern-1.9ex{#1}}

\def\h{\hat}
\def\b{\bar}

\def\={\,\hat{=}\,}
\def\f{\frac}

\def\Lie{\mathcal{L}}
\def\ps{\boldsymbol{\Gamma}_{\rm cov}}
\def\psR{\boldsymbol{\Gamma}_{\!\scriptscriptstyle{\cal R}}}
\def\psRh{\boldsymbol{\Gamma}_{\!\scriptscriptstyle{\cal {\hat R}}}}
\def\psDelta{\boldsymbol{\Gamma}_{\Delta}}
\def\psrad{{\boldsymbol{\Gamma}}_{\rm rad}}
\def\psRDelta{{\boldsymbol\Gamma}_{\mathcal{R}[\Delta]}}
\def\psSigma{\boldsymbol{\Gamma}_{{}_{\!\Sigma}}}
\def\bfomega{\boldsymbol{\omega}}
\def\bfomegarad{{\boldsymbol{\omega}}_{{}_{\rm rad}}}
\def\bfomegaR{\boldsymbol{\omega}_{\scriptscriptstyle{\cal R}}}
\def\bfomegaRh{\boldsymbol{\omega}_{\scriptscriptstyle{\cal {\hat R}}}}
\def\bfomegaSigma{\boldsymbol{\omega}_{{}_{\Sigma}}}

\def\hM{\hat{M}}
\def\hg{\hat{g}}
\def\hR{\hat{R}}

\def\hD{\hat{D}}
\def\hn{\hat{n}}
\def\hq{\hat{q}}

\def\G{\mathfrak{G}}

\def\B{\mathfrak{B}} 
\def\T{\mathfrak{T}}
\def\J{\mathfrak{J}}

\def\super{\hat{\mathfrak{s}}}
\def\tran{\hat{\mathfrak{t}}}

\def\qo{\mathring{q}}

\def\vo{\mathring{v}}

\def\betao{\mathring{\beta}}
\def\chio{\mathring{\chi}}
\def\varpio{\mathring{\varpi}}
\def\psio{\mathring{\psi}}
\def\alphao{\mathring\alpha}

\def\ello{\mathring{\ell}}
\def\no{\mathring{n}}

\def\Do{\mathring{D}}
\def\omegao{\mathring{\omega}}
\def\epsilono{\mathring{\epsilon}}

\def\WIH{\mathfrak{h}}

\def\psio{\mathring{\psi}}

\def\rmd{\mathrm{d}}

\def\S{\mathcal{S}}

\def\T{\mathcal{T}}
\def\Q{\mathcal{Q}}

\def\DR{\mathcal{D}\!_{{}_{\R}}}
\def\DRh{\mathcal{D}_{\scriptscriptstyle{\cal {\hat R}}}}
\def\DRhp{\mathcal{D'}_{\!\!\scriptscriptstyle{\cal {\hat R}}}}

\def\R{\mathcal{R}}
\def\F{\mathcal{F}}
\def\Fxi{\hat{\mathcal{F}}^{(\xi)}}

\def\Rh{\hat{\mathcal{R}}}

\def\vac{\hat{\mathfrak{D}}}
\def\co{\mathring{c}}

\def\nn{\nonumber}

\newcommand{\na}{\nabla}

\renewcommand{\d}{\delta}  \newcommand{\eps}{\epsilon} 
   \newcommand{\vth}{\vartheta}

 \newcommand{\s}{\sigma}      
  \let\Th=\Theta  
 \let\Om=\Omega

\newcommand{\sscr}{\scriptscriptstyle\rm}

\newcommand{\eqons}{\,\doteq\,}

\usepackage{pgf}
\makeatletter
\newcommand*{\pgfunderleftarrow}{%
  \@ifstar
    {\let\ifpgf@depth\iftrue\mathpalette\@pgfunderleftarrow}
    {\let\ifpgf@depth\iffalse\mathpalette\@pgfunderleftarrow}%
}
\newcommand*{\@pgfunderleftarrow}[2]{%
  #2%
  \edef\pgf@math@fam{\the\fam}%
  \pgfpicture
    \pgfsetbaseline{0pt}
    \pgf@relevantforpicturesizefalse      
    \pgfsetroundcap                       
    \pgfsetarrowsend{to}
    \pgfutil@tempdima=0.28pt%
    \advance\pgfutil@tempdima by.8\pgflinewidth%
    \pgfutil@tempdima-4\pgfutil@tempdima
    \sbox\pgfutil@tempboxa{$\m@th\fam\pgf@math@fam#1#2$}%
    \advance\pgfutil@tempdima-\dp\pgfutil@tempboxa
    \pgfutil@tempdimb\wd\pgfutil@tempboxa
    \pgfpathmoveto{\pgfqpoint{0pt}{\pgfutil@tempdima}}%
    \pgfpathlineto{\pgfqpoint{-\pgfutil@tempdimb}{\pgfutil@tempdima}}%
    \pgfusepath{stroke}
    \ifpgf@depth
      \pgf@relevantforpicturesizetrue
      \pgfpathmoveto{\pgfqpoint{0pt}{-\pgfutil@tempdimb}}%
      \pgfusepath{use as bounding box}%
    \fi
  \endpgfpicture
}
\makeatother
\newcommand{\pbi}[1]{\pgfunderleftarrow{#1}}

\usepackage{mathrsfs}

\begin{document}

\title{Null Infinity and Horizons: \\ A New Approach to Fluxes and Charges}

\author{Abhay Ashtekar}
\email{ashtekar.gravity@gmail.com}
\affiliation{Institute for Gravitation and the Cosmos, Pennsylvania State 
University, University Park, PA 16802, USA}
\affiliation{Perimeter Institute for Theoretical Physics, 31 Caroline St N, Waterloo, ON N2L 2Y5, Canada}

\author{Simone Speziale}
\email{simone.speziale@cpt.univ-mrs.fr}
\affiliation{Aix Marseille Univ., Univ. de Toulon, CNRS, CPT, UMR 7332, 13288 Marseille, France}

\begin{abstract}

\medskip

\noindent We introduce a Hamiltonian framework tailored to degrees of freedom (DOF) of field theories that reside in suitable 3-dimensional open regions, and then apply it to the gravitational DOF of general relativity. Specifically, these DOF now refer to open regions $\Rh$ of null infinity $\scrip$, and $R$ of black hole (and cosmological) horizons $\Delta$ representing equilibrium situations. At $\scrip$ the new Hamiltonian framework yields the well-known BMS fluxes and charges \cite{gerochrev,aams,Dray1984,Dray1985}. By contrast, all fluxes vanish identically at $\Delta$ just as one would physically expect \cite{akkl2}. In the companion paper \cite{aass1} we showed that, somewhat surprisingly, the geometry and symmetries of $\scrip$ and $\Delta$ descend from a common framework. This paper reinforces that theme: Very different physics emerges in the two cases from a common Hamiltonian framework because of the  difference in the nature of degrees of freedom on $\scrip$ and $\Delta$, discussed in \cite{aass1}. Finally, we compare and contrast this Hamiltonian approach with those available in the literature.

\end{abstract}

\maketitle

\tableofcontents

\section{Introduction}
\label{s1}

In a companion paper \cite{aass1}, we showed that the notion of Weakly Isolated Horizons (WIHs) $\WIH$ provides a common platform to discuss both black hole horizons $\Delta$ in equilibrium, \emph{as well as} null infinity $\scri$. In particular, the familiar geometrical structures and the BMS group of symmetries at $\scri$ can be systematically recovered from the WIH framework. This seems surprising at first since WIHs are generally associated with the black hole horizons $\Delta$ that lie in the strong curvature region and there is no flux of gravitational radiation (or matter fields) across them. Therefore WIHs have had the connotation of representing geometries of boundaries that are in equilibrium. $\scri$ on the other hand lies in the weak field, asymptotic regions of space-time and provides the arena to analyze the physics of gravitational waves (and zero rest mass fields). We showed that these striking differences in the physics arises from the fact that while $\Delta$ is a WIH in physical space-times $(M, g_{ab})$ where $g_{ab}$ satisfies Einstein's equations, $\scri$ is a WIH in the conformally completed space-time $(\hM,\hg_{ab})$ where $\hg_{ab}$ satisfies the conformal Einstein's equations. 

The common underlying WIH description also allows one to understand the symmetries of $\scri$ and $\Delta$ using a single framework \cite{aass1}. Following Noether, the boundary symmetries should be related to generators of canonical transformations on the phase space. However this relation is far more subtle in presence of `leaky boundaries' such as $\scrip$ and one has to carefully resolve potential ambiguities in the definitions. There are various approaches to address these issues, most notably the Wald-Zoupas \cite{wz} and Barnich-Brandt prescriptions \cite{Barnich:2001jy}. In this paper we introduce a new approach in which fluxes associated with symmetries arise as Hamiltonians generating the action of BMS symmetries on the radiative phase space. This procedure gives conceptual precedence to fluxes; charges are obtained by a systematic `integration' of these fluxes. We will find that very different physics arises at null infinity $\scrip$ and black hole horizons $\Delta$  from a common framework for the same reason as in Ref. \cite{aass1}: $\Delta$ is a WIH with respect to $(M, g_{ab})$, while $\scri$ is a WIH with respect to $(\hM,\hg_{ab})$. For definiteness we will focus on $\scrip$ since physics at $\scrip$ is much more interesting than that at $\scrim$ for isolated gravitating systems of physical interest. But all our considerations apply to $\scrim$ as well. Similarly, even though we generally refer to $\Delta$ as black hole horizons in equilibrium, all our results continue to hold if $\Delta$ is a cosmological WIH.

In section \ref{s2}, we introduce the new Hamiltonian framework using a familiar example: a free scalar field in Minkowski space-time. The strategy is to extract phase spaces $\psR$ adapted to 3-dimensional open regions $\R$ that represent `degrees of freedom of the field that reside in $\R$'. In section \ref{s3.1} we introduce phase spaces $\psR$ associated with regions $\R$ of a Cauchy surface and, in section \ref{s3.2}, phase spaces $\psRh$ associated with regions  $\Rh$ of $\scrip$. We show that this framework yields the `physically correct'  BMS fluxes, i.e. fluxes that agree with those obtained by using the stress-energy tensor of the scalar field without, however, having to refer to the stress-energy tensor anywhere. Therefore, the strategy can be carried over the gravitational field of general relativity, for which there is no stress-energy. 

In sections \ref{s3} and \ref{s4} we apply the strategy (rooted in \cite{aams}) to the gravitational case. We begin with the covariant phase space $\ps$ of vacuum%
\footnote{As usual, the restriction to vacuum equations is made to avoid the introduction of specific matter fields and the corresponding phase spaces. The analysis can be extended to allow zero rest mass fields as sources. In the discussion of $\scrip$ one has to further restrict oneself to the $\Lambda=0$ case because, to qualify as a WIH, $\scrip$ has to be null.}
solutions to Einstein's equations \cite{aaam-cmp,cw,abr} that admit a WIH $\WIH$ as a boundary, which could be either $\scrip$ or $\Delta$. From the full $\ps$ one can extract the phase spaces $\psRh$ and $\psR$, adapted to open regions $\Rh$ and $\R$ of $\scrip$ and $\Delta$, and use Hamiltonian methods to obtain expressions of fluxes $F_\xi [\Rh]$ and $F_\xi[\R]$, associated with symmetry vector fields $\xi^a$, across $\Rh$ and $\R$. Although the conceptual steps are the same, the flux expressions are dramatically different because $\psRh$ and $\psR$ inherit quite different symplectic currents from $\ps$. As on $\ps$, the symplectic current on $\Gamma_{\WIH}$ is a 3-form, but now pulled-back to $\WIH$. If $\WIH$ is a black hole horizon, the pull-back of the symplectic current to $\Delta$ simply vanishes. On the other hand, at $\scrip$, the terms that would have vanished are multiplied by inverse powers of the conformal factor $\Omega$ relating $g_{ab}$ to $\hg_{ab}$, and the pull-back is non-zero. As a direct consequence, while fluxes $F_\xi [\R]$ vanish on any open region $\R$ of $\Delta$,\, $F_\xi [\Rh]$ provide non-trivial observables on the phase spaces $\psRh$ that capture the radiative degrees of freedom at $\scrip$. A notable feature of this common procedure for $\Delta$ and $\scrip$ is that one works just with the symplectic structure. This is in contrast to other approaches in the literature (see, e.g., \cite{PhysRev.150.1039,Geroch:1981ut,wz,cfp}), where one has to introduce additional inputs such as  a `linkage' or  a `preferred' symplectic potential. In asymptotically de Sitter space-times, for example, the preferred symplectic potential satisfying all the requirements spelled out in \cite{wz,cfp} does not exist \cite{kl2021} but one can use the Hamiltonian framework introduced in this paper to arrive at physically viable fluxes \cite{akl-lambda}.

Finally, as noted above, the 2-sphere `charge aspects' $\Q^{(\xi)}_{ab}$ at $\scrip$ are obtained by `integrating' flux aspects $\Fxi_{abc}$ using conformal Einstein's equations. This step requires the full set of field equations --not just their pull-backs to 3-surfaces  $\Rh$-- because while fluxes at $\scrip$ refer only to the radiative degrees of freedom, charges refer also to the Coulombic ones. In other words, one has to step outside the phase spaces $\psRh$ and $\psR$, that know only about fields and equations intrinsic to $\Rh$ and $\R$, and return to the full covariant phase space $\ps$ which accesses full Einstein's equations, including those that link the Coulombic and radiative aspects. 

Section~\ref{s5} summarizes the main results and briefly discusses some applications of this framework. In particular, we point out that the common platform that unifies the discussion of null infinity $\scrip$ and black hole horizons $\Delta$ is being used in a number of contexts both classical and quantum gravity. We also provide a short overview of the conceptual similarities and differences between our framework and those available in the literature. Details of this comparison --particularly with the approach pioneered by Wald and Zoupas \cite{wz} and extended by Chandrasekharan, Flanagan and Prabhu \cite{cfp} and others-- are presented in Appendix \ref{a1}.

Our conventions are as follows.  In the discussion of black hole horizons $\Delta$, the underlying physical space-times is denoted by $(M, g_{ab})$. The torsion-free derivative operator compatible with $g_{ab}$ is denoted by $\nabla$ and its curvature tensors are defined via:  $2\nabla_{[a}\nabla_{b]} v_c = R_{abc}{}^d v_d$,\, $R_{ac} = R_{abc}{}^b$, and $R=g^{ab} R_{ab}$. The Penrose conformal completions used in the discussion of $\scrip$ is denoted by  $(\hM, \hg_{ab})$ and the corresponding derivative operator and curvature tensors carry a hat. Null normals to $\Delta$ and $\scrip$ are assumed to be future directed. All fields are assumed to be smooth for simplicity but this requirement can be weakened substantially (in particular to allow for the possibility that the Newman-Penrose curvature component $\Psi_1^\circ$ may violate peeling). \emph{If there is a possibility of ambiguity, we will use $\=$ to denote equality that holds only at $\scrip$ or $\Delta$.}  

\section{Phase spaces $\psR$ associated with finite regions $\R$}
\label{s2}

Our goal is to arrive at fluxes and charges associated with symmetries of $\scrip$ and $\Delta$ using a Hamiltonian framework that is tailored to degrees of freedom that reside in their finite sub-regions $\Rh$ and $\R$. This procedure involves several new elements. Therefore, in this section we will make a detour and illustrate the strategy and main constructions using a simple and familiar example: a scalar field in Minkowski space-time. We will then apply this procedure to $\scrip$ in section \ref{s3} and to $\Delta$ in section \ref{s4}.

Recall that there is a well-known covariant phase space framework for classical field theories in globally hyperbolic space-times \cite{cm,aaam-cmp,cw,abr,wz} (whose foundations can be traced back all the way to Lagrange!). It provides a natural platform to define Hamiltonians generating canonical transformations associated with symmetries. The covariant phase space $\ps$ consists of suitably regular solutions of field equations. The idea now is to extract from $\ps$, a phase space $\psR$ associated with certain 3-dimensional regions $\R$ in the space-time under consideration. Intuitively, $\psR$ can be thought of as capturing the degrees of freedom of the field that are registered on $\R$. We will show that the Hamiltonians on $\psR$  have the physical interpretation of fluxes associated with symmetries: they agree with the standard expressions of fluxes obtained from the stress-energy tensor of the field and symmetry vector fields.

This section is divided into two parts. In the first we explain the main ideas using finite regions with Cauchy surfaces in Minkowski space and in the second we extend them to obtain fluxes of BMS momenta across finite regions of $\scrip$. This approach uses some basic properties of infinite dimensional linear topological spaces. We will spell out the details since these structures are not commonly used in the physics literature on systems with boundaries. In the second part we discuss phase spaces $\psRh$ associated with finite regions $\Rh$ of $\scrip$.

\subsection{Example: Klein Gordon Field in Minkowski Space}
\label{s4.1}

Consider a scalar field $\phi$ in Minkowski space satisfying the Klein-Gordon equation $\Box\, \phi -\mu^2\phi =0$. The covariant phase space $\ps$ of this system consists of (suitably regular) solutions $\phi$ to this equation \cite{cm}. Since $\ps$ is a vector space, a tangent vector $\delta$ at any $\phi \in \ps$ is also represented by a solution to the KG equation which we will denote by $\delta\phi$.\, $\ps$ is naturally equipped with a (weakly non-degenerate) symplectic structure $\bfomega$,\, a 2-form on $\ps$ whose action on tangent vectors $\delta_1, \delta_2$ at any $\phi$ is given by
\footnote{Recall that $\bfomega$ is said to be weakly non-degenerate if $\bfomega(\delta_1, \delta_2) =0$ for all $\delta_2$, if and only if $\delta_1 =0$. This property does not imply that $\bfomega$ admits an inverse on $\ps$. But weak non-degeneracy suffices to discuss Hamiltonians generating infinitesimal canonical transformations.}   
\be \label{symp1} \bfomega\!\mid_{{}_\phi} (\delta_1, \delta_2)= \int_\Sigma \big((\delta_1 \phi) \nabla_a (\delta_2 \phi)\, - \, (\delta_2 \phi) \nabla_a (\delta_1 \phi)\big)\, \epsilon^a{}_{bcd}\,\, \rmd S^{bcd}\, \equiv \, 
\int_\Sigma \J_{bcd} \,\rmd S^{bcd}\, , \ee
where $\Sigma$ is a Cauchy surface. The 3-form $\J_{bcd}$ is referred to as the symplectic current. As is well known, for any Killing field $\xi^a$,\, $\delta_\xi := \Lie_\xi \phi$ is a tangent vector at a point $\phi\in \ps$ and, given any other tangent vector $\delta$, using the fall-off conditions used in the definition of $\ps$ one finds that there is a function $H_\xi(\phi)$ on $\ps$ satisfying
\be \label{KGflux1} \bfomega\!\mid_{{}_\phi} \!(\delta_\xi, \delta) = \delta H_\xi\!\mid_{{}_\phi}\,  \ee 
for all $\phi \in \ps$. Thus, the vector field $\delta_{\xi}$ on $\ps$ is the infinitesimal canonical transformation generated by $H_\xi(\phi)$ \cite{cm}. $H_\xi$ is unique up to an additive constant which one fixes by demanding that it vanish at the solution $\phi =0$. The physical interpretation of this $H_\xi$ comes from the fact that it equals
the conserved flux $F_\xi$ across $\Sigma$ constructed from the stress-energy tensor $T_{ab} = \nabla_a \phi \nabla_b \phi - \f{1}{2}\, g_{ab}\, (\nabla_c\phi\ \nabla^c\phi)$ of the Klein-Gordon field and the given Killing field $\xi^a$: 
\be \label{KGflux2} H_\xi  =  F_\xi  :=\int_\Sigma T_{ma}\, \xi^m \epsilon^a{}_{bcd}\, \rmd S^{bcd}\, .\ee
for all solutions $\phi \in \ps$. If for example, $\xi^a$ is a unit time translation, $H_\xi$ equals the energy flux $E$ across $\Sigma$ and, since $\Sigma$ is a Cauchy surface, $E$ is also the total energy in the field. Thus, the Hamiltonian $H_\xi$ of the phase space framework provides us with the physically correct expressions of fluxes even though the procedure that led us to $H_\xi$ has no direct knowledge of $T_{ab}$.
We will now show that this interplay between the Hamiltonian framework of $\ps$ and stress-energy tensor can be extended to phase spaces $\psR$ associated with certain 3-dimensional finite regions $\mathcal{R}$.

Let us then consider an open region $\R$ in $\Sigma$ with compact closure.  The flux $F_\xi [\R]$ across $\R$   --constructed from $T_{ab}$-- is simply the restriction of the integral in (\ref{KGflux2}) to $\R$. A natural question is if this flux can also be obtained as a Hamiltonian in a phase space framework that only knows about $\R$. We will first show that the answer is in the affirmative and then transport these ideas to null infinity in section \ref{s2.2} to calculate fluxes across local regions of $\scrip$. 

For simplicity let us assume that $\Sigma$ is a 3-plane (extension to a general $\Sigma$ is straightforward). The first step is to extract a local phase space $\psR$ from $\ps$ by focusing on \emph{degrees of freedom in $\phi$ that are intrinsic to $\R$}. These are encoded in the initial data 
$(\varphi :=\phi\!\mid_{{}_{\R}}, \, \pi := \dot\phi\!\mid_{{}_{\R}})$ of $\phi$, restricted to the open set $\R$. For brevity, we will denote a point in $\psR$ by $\gamma \equiv (\varphi,\, \pi)$. 
In the discussion of $\ps$ above, we did not spell out the details of regularity conditions because there are well-known choices (see, e.g., \cite{cm}). Since local phase spaces are not discussed in the literature, let us spell out our choice. $\psR$ will consist of pairs $(\varphi, \pi)$ on $\R$ such that
\be \label{energynorm2}  ||\gamma||^2_{{}_\R}\,  :=  \int_{\R} \big(\pi^2 + D_a\varphi D^a\varphi +\mu^2 \varphi^2 \big)\, \rmd^3 x\,  <\infty \, .\ee
That is, pairs $(\varphi, \pi)_\R$ that belong to $\psR$ constitute $H^1 (\R) \oplus L^2(\R)$, where $H^1(\R)$ is the first Sobolev space. 
Physically, the norm is simply the energy-flux of the initial data restricted to $\R$. However, its actual value will not play any role for our purposes. We will only need the topology this norm induces on $\psR$ (see Remark 5 below).

The symplectic structure $\bfomegaR$ on $\psR$ is obtained by just restricting the integral in (\ref{symp1}) to $\R$:

\be \label{symp2} \bfomegaR\!\!\mid_{{}_\gamma} \!(\delta_1, \delta_2)\, =\, \,\int_{\R} \J_{abc}\, \rmd S^{abc}\, =\,
\int_{\R} \big(\delta_1 \varphi \, \delta_2 \pi\, - \, \delta_2 \varphi \, \delta_1 \pi\big)\, \rmd^3x\, .\ee
It is again weakly non-degenerate, and also continuous in the topology induced by the norm (\ref{energynorm2}). Let us consider a Killing field $\xi^a$ that is tangential to $\Sigma$ and ask if the vector field $\delta_\xi$ represents an infinitesimal canonical transformation also on $\psR$. It suffices to evaluate $\bfomegaR{\mid_{{}_\gamma}} (\delta_\xi, \delta)$ where $\delta_\xi \equiv (\Lie_\xi \varphi,\, \Lie_\xi \pi)$, and $\delta$ are arbitrary constant vector fields on the \emph{vector space} $\psR$. The domain of definition of $\delta_\xi$ is the dense subspace $\DR$ of $\psR$ on which $(\Lie_\xi \varphi,\, \Lie_\xi \pi)$ also belongs to $\psR$. 
On this subspace we have

\be \label{KGflux3} \bfomega\!\mid_{{}_\gamma} \!(\delta_\xi, \delta) = \delta \int_\R \big(\Lie_\xi \varphi)\,\pi\, \rmd^3 x \,-\, \oint_{\partial\R} \!(\pi\, \delta\varphi)\, \xi^a dS_a\, . \ee
Thus, if $\xi^a$ happens to be tangential to $\partial\R$ \, (e.g., if it is a rotation and $\partial\R$ is spherical),\, then, $\delta_\xi$ is indeed a Hamiltonian vector field on $\DR$, generated by $H_\xi [\R] = \int_\R \big(\Lie_\xi \varphi)\,\pi \,\rmd^3 x$. (On any phase space, there is freedom to add a constant $C_\xi$ to the Hamiltonian. As usual, it has been eliminated by requiring that $H_\xi$ should vanish at the point ($\varphi=0,\, \pi=0$).) By inspection $H_\xi [\R]$ is continuous on $\DR$ and therefore admits unique continuous extension to all of $\ps$. Furthermore, it equals the flux $\F_\xi [\R] = \int_\R T_{ma}\, \xi^m\, \epsilon^{a}{}_{bcd} \rmd S^{bcd}$, defined by the stress-energy tensor on $\psR$. Finally, the infinitesimal canonical transformation it generates can be exponentiated --the 1-parameter family of finite canonical transformations is just the action of the diffeomorphism group generated by $\xi^a$ on $\psR$.

What if $\xi^a$ is not tangential to $\partial\R$ (e.g., a space-translation)? Then we restrict ourselves to the subspace ${\DR}^{\!\!\prime}$ of $\DR$ on which $\pi\!\!\mid_{{}_{\partial\R}} =0$, which is also dense in $\psR$. On this subspace the surface term on the right hand side of (\ref{KGflux3}) vanishes and we again have a Hamiltonian vector field generated by $H_\xi[\R]$.
As before, $H_\xi[\R]$ admits a unique continuous extension to all of $\psR$ 
\be \label{H1} H_\xi[\R] := \int_\R \big(\Lie_\xi \varphi)\,\pi \rmd^3 x\, , \ee 
that agrees with the flux $F_\xi[\R]$ given by $T_{ab}$:\, $H_\xi[\R] = F_\xi[\R]$ on all of $\psR$. Note, however, that in this case the infinitesimal canonical transformation cannot be exponentiated although the Hamiltonian admits a continuous extension to $\psR$. (This is analogous to a rather common occurrence in quantum mechanics, e.g. for a particle on half line and the translation generator $-i\hbar\,\rmd /\rmd x$.) In the next sub-section we will extend these results to subregions $\R$ of $\scrip$.\\

\emph{Remarks:} 

1. Note that our considerations go through for any smooth, divergence-free vector field  $\xi^a$ that is tangential to $\R$ --it need not be a Killing field. (If $D_a\xi^a \not=0$, the required integration by parts would yield an additional contribution to the volume term and\, $\bfomega\!\mid_{{}_\gamma} \!\!(\delta_\xi, \delta)$\, would be no longer of the form $\delta H_\xi$ on a dense subspace.) Its action on the $(\varphi, \pi)$ pairs in $\psR$ is again a well-defined infinitesimal canonical transformation with the Hamiltonian given by $H_\xi [\R]$ of Eq.~(\ref{H1}), which again agrees with the flux $F_\xi [R]$ determined by the stress-energy tensor and the vector field $\xi^a$. But these integrals do not have a simple physical interpretation since they do not descend form observables naturally defined on the full phase space $\ps$.
 
2. There is another, more interesting class of vector fields $\xi^a$ one can consider: Killing fields that are transverse --rather than tangential-- to $\R$, such as  time-translations and boosts. One can readily extend the strategy to include these vector fields. However in these cases the phase space vector fields $\delta_\xi$ on $\psR$ are not expressible as Lie derivatives of $(\varphi,\pi)$ along $\xi^a$; one to first define them using the field equations satisfied by $\phi$ and the relation between $\ps$ and $\psR$. (For example for the unit time translation vector field $\xi^a$ orthogonal to $\R$, we have $\delta_\xi (\varphi, \pi) = (\pi,\, (D^2-\mu^2) \varphi)$.) Then the procedure can be repeated and the resulting $H_\xi$ agrees with the flux $F_\xi$. We focused on symmetry vector fields $\xi^a$ that are tangential to the 3-d region $\R$ because our main motivation comes from $\scrip$ and $\Delta$. They are 3-dimensional and symmetry vector fields are tangential to them.

3. The right side of (\ref{KGflux3}) can also be written as
\be  \label{KGflux4} \bfomega\!\mid_\gamma \!(\delta_\xi, \delta) = - \delta \int_\R \big(\Lie_\xi \pi)\, \varphi\, \rmd^3 x \,-\, \oint_{\partial\R} \!(\varphi\, \delta\pi)\, \xi^a dS_a\,  \ee
and the surface term vanishes on a dense subspace $\DR^{\,\,\prime\prime}$ consisting of pairs $(\varphi, \pi)$ in which $\pi$ is $C^1$ and $\phi$ vanishes on $\partial \R$. Therefore one may be tempted to say that $\int_\R \big(\Lie_\xi \pi)\, \varphi\big)\,\rmd^3 x$ is also a candidate Hamiltonian generating the infinitesimal canonical transformation $\delta_\xi$ on $\psR$. However, this is incorrect because, since this candidate Hamiltonian involves \emph{derivatives} of $\pi$, it cannot be extended continuously to full $\psR$. That is, since the energy norm (\ref{energynorm2}) that endows $\psR$ with its topology contains only the Klein-Gordon field $\phi$ and its first derivatives --i.e., only $\varphi,\, D_a\varphi$, and $\pi$-- the new candidate for $H_\xi$ fails to be a continuous function on $\DR^{\,\,\prime\prime}$, whence it cannot be extended to $\psR$. Thus, the Hamiltonian on $\psR$ is unambiguous.

4. We want to emphasize, however, that this discussion does \emph{not} imply that all subtleties involving boundary terms are irrelevant. For example, they cannot be circumvented in gauge theories simply by using topological arguments on suitable function spaces. In particular, the symplectic structure on $\ps$ is gauge invariant because of the boundary conditions imposed at infinity, and would typically cease to have this property on $\psR$. The interplay between topological arguments and subtleties associated with gauge invariance is interesting in its own right. But it would be a digression to discuss these issues here because our main interest is on $\scrip$ and, as we will discuss in section \ref{s3} one can isolate the gauge invariant degrees of freedom on  $\scrip$.

5. It is clear from the main discussion of this subsection that the norm (\ref{energynorm2}) serves only to induce the topology: it implies that a sequence of points $(\varphi_n, \pi_n)_\R$ in $\psR$ converge to $(\varphi_\circ,\pi_\circ)$\, if \, $\varphi_n,\, \pi_n$ and $D_a\varphi_n$ converge to $\varphi_\circ,\, \pi_\circ$\, and\, $D_a\varphi_\circ$ in the $L^2$-sense. The precise value of this norm itself is not relevant; only the topology it induces is used. This fact will be useful in the next subsection. Finally, note that while we  introduced the phase space $\psR$ associated with finite regions $\R$ to illustrate the strategy we will use at  WIHs, the idea of using phase spaces associated with finite 3-d regions has other applications as well since --e.g. at the future space-like infinity of asymptotically de Sitter space-times-- since they encode the degrees of freedom just in that region.

\subsection{Scalar field at null infinity} 
\label{s2.2}

To extend these considerations to null infinity, let us now consider a massless scalar field satisfying the wave equation\, $\Box\, \phi =0$\, on Minkowski space $(M, g_{ab})$. Let $(\h{M}, \h{g}_{ab})$ be any conformal completion in which $\scrip$ is divergence-free, i.e., its null normal $\hn^a$ satisfies $\h\nabla_a\hn^a \= 0$. Given a $\phi \in \ps$, the conformally rescaled field $\h\phi := \Omega^{-1} \h\phi$ is well-defined at $\scrip$. By pulling back the symplectic current to $\scrip$, the symplectic structure $\bfomega$ on $\ps$ can be rewritten as:
\ba \label{symp3} \bfomega\!\mid_\phi (\delta_1, \delta_2) &=& \int_{\scrip} \big((\delta_1 \h\phi) \h\nabla_a (\delta_2 \h\phi)\, - \, (\delta_2 \h\phi) \h\nabla_a (\delta_1 \phi)\big)\,\, \h\epsilon^a{}_{bcd}\,\, \rmd \h{S}^{bcd}\, 
\nonumber\\
&\equiv& \, \int_{\scrip} \h{\J}_{bcd} \,\,\rmd \h{S}^{bcd}\, . \ea

Let us now consider an open region $\Rh$ of $\scrip$, with topology $\mathbb{S}^2\times \mathbb{R}$, bounded by any two cross-sections $S_1$ and $S_2$. Since the phase space $\psRh$ is to consist of degrees of freedom that reside in $\Rh$, we are led to consider restrictions of $\h\phi$ to $\Rh$ and introduce a suitable topology on these fields. The idea again is to introduce this topology by choosing a norm which ensures that two  elements of $\psR$ are close to one another if they and their first derivatives are close in the $L^2$ sense. Therefore we will take the norm to be
\be \label{norm1} ||\h\phi||^2_{{}_{\R}}\, =\,  \int_{\R} \big(|\h{n}^a D_a \h\phi|^2 + \hq^{ab}\, \hD_a \h\phi\, \hD_b \h\phi + \f{1}{l^2} |\h\phi|^2\big)\, \rmd^3\! \scrip \ee 
where $l$ is a constant with dimensions of length, $\hq^{ab}$ is an inverse of $\hq_{ab}$ (i.e., a field satisfying $\hq^{cd}\, \hq_{ac}\, \hq_{bd} = \hq_{ab}$), and $\rmd^3 \scrip$ the natural volume element on $\scrip$ in the conformal frame $(\hq_{ab}, \hn^a)$. This norm depends on the auxiliary structures --a specific conformal frame, inverse metric $\hq^{ab}$, and the constant $l$-- we introduced on $\scrip$, but the topology it induces is insensitive to these choices. The symplectic structure is given by restricting the integral in (\ref{symp3}) to $\Rh$:\, $\bfomegaRh\!\!\mid_{{}_\gamma} (\delta_1, \delta_2)\, =\,\int_{\Rh} \h{\J}_{abc}\,\, \rmd \h{S}^{bcd}$. Again, it is weakly non-degenerate and continuous on $\psRh$.

Next, recall that the BMS vector fields $\h\xi^a$ satisfy $\Lie_\xi \hq_{ab} = 2\beta \hq_{ab}$ and $\Lie_\xi \hn^a = -\beta\, \hn^a$. 
Tangent vectors $\delta_\xi$ agin correspond to the fields $\delta_\xi \h\phi \equiv \Lie_\xi \h\phi$. However, since the scalar fields $\h\phi$ have conformal weight $-1$, the action of the Lie derivative has an extra term relative to its action on functions with zero conformal weight:
\be \label{modification} \Lie_\xi \,\,\h\phi = \xi^a D_a \h\phi + \beta\, \h\phi \, .\ee
(For the supertranslation vector fields $\xi^a = \super\,\hn^a$, we have $\Lie_{\super\hn}\,\, \hq_{ab} = 0$, i.e., 
$\beta =0$, whence $\Lie_{\super\,\hn}\, \h\phi = \super\,\hn^a D_a \h\phi$.)\, The second term in the right side of Eq.~(\ref{modification}) ensures that $\delta_\xi \h\phi$ is again a scalar field with conformal weight $-1$.  Hence the infinitesimal transformation generated by $\xi^a$ is well-defined on the dense subspace $\DR$ of $\psR$ on which $\Lie_\xi \h\phi$ is again in $\psR$. Given any constant vector field $\delta$ on $\psR$  and $\h\phi \in \DRh$, a simple calculation yields:
\be \bfomegaRh\!\!\mid_{{}_{\h\phi}} (\delta_\xi, \delta) = \int_{\Rh}\! \big[(\Lie_{\xi} \h\phi) (\hn^a D_a\delta\h\phi)\, +\, (\Lie_\xi \delta\h\phi)(\hn^a D_a\h\phi) \big]\, \rmd^3\! \scrip 
+ \oint_{\partial\Rh}\!\! \big[\delta\h\phi\,(\hn^a D_a \h\phi)\,\xi^m\big] \rmd \h{S}_m \, .\ee
Therefore on the subspace ${\DRh}^{\!\!\prime}$ of $\DRh$ on which $\hn^a D_a \phi$ vanishes on the boundary $\partial\R$, we again have:
\be \bfomegaRh\!\!\mid_{{}_{\h\phi}} (\delta_\xi, \delta) = \delta \int_{\Rh} (\Lie_\xi \phi)(\Lie_{\hn} \h\phi)\, \rmd^3\! \scrip 
\,\equiv \, \delta H_\xi (\phi)
\ee
As before, since ${\DRh}^{\!\!\prime}$ is also dense in $\psR$ and the function $H_\xi$ is continuous, it admits a unique continuous extension to the full phase space $\psR$. In this sense, $\delta_\xi$ is again an infinitesimal canonical transformation generated by the Hamiltonian $H_\xi (\phi)$ on $\psRh$. The Hamiltonian $H_\xi$ represents the flux of the BMS momentum across the portion $\Rh$ of $\scrip$. Again, if $\xi^a$ happens to be tangential to the boundary, the Hamiltonian vector field $\delta_\xi$ generates a 1-parameter family of  \emph{finite} canonical transformations; for more general vector fields we only have the infinitesimal canonical transformations. 

Let us summarize. $\psRh$ captures the degrees of freedom $\h\phi\!\mid_{\Rh}$ in the scalar field $\phi$ that are registered in the region $\Rh$ of $\scrip$. In the literature, $\hat\phi\!\mid_{\scrip}$ is often referred to as the \emph{radiation field} because it is the coefficient of the `$1/r$-part' in the asymptotic expansion of $\phi$ in Bondi-type coordinates $u,r,\vartheta,\varphi$. Thus $\psRh$ is the space of radiative degrees of freedom that reside in $\Rh$. The topology on $\psRh$ ensures that a sequence $\h\phi_n $ converges to $\h\phi_\circ$ as $n \to \infty$ if and only if $\h\phi_n$ and their first derivatives $D_a \h\phi_n$ converge to $\h\phi_\circ$ and $D_a\h\phi_\circ$ in the $L^2$ sense. The symplectic structure $\bfomegaRh$ on $\psRh$ (obtained by restricting $\bfomega$ on $\ps$ to the degrees of freedom captured in $\psRh$) is weakly non-degenerate and is  continuous on $\psRh$. The action of every BMS vector field $\xi^a$ on $\scrip$ induces a vector field $\delta_\xi \h\phi$ on a dense subspace of $\psRh$ that preserves $\bfomegaRh$. This infinitesimal canonical transformation is generated by a unique  Hamiltonian $H_\xi$ that is continuous on the full phase space $\psRh$ and vanishes at $\h\phi =0$. Note that the topology on $\psRh$ is precisely such that $\bfomegaRh$ and $H_\xi$ are continuous on $\psRh$. And then the Hamiltonian $H_\xi$ agrees with the flux $F_{\xi}$ of the BMS momentum across $\Rh$, computed using the stress-energy tensor on all of $\psRh$.

\section{General Relativity: Gravitational Field at Null Infinity}
\label{s3}

{Let us now turn to the gravitational field in full, non-linear general relativity and consider space-times that admit a WIH boundary. In this section we will focus on the case when the WIH is $\scrip$ and show that the Hamiltonian methods introduced in section \ref{s2} lead to the standard  expressions for fluxes of BMS momenta across regions $\Rh$ of $\scrip$. In section \ref{s4}, we will show that the same procedure implies that the fluxes associated with symmetries of black hole WIHs $\Delta$ vanish identically. 

We begin in section \ref{s3.1} with asymptotically flat space-times with a complete $\scrip$. We first recall from Refs.~\cite{aa-radiativemodes,aaam-cmp,aams,ashtekar1987asymptotic} how degrees of freedom can be isolated at $\scrip$ to obtain a radiative phase-space $\psrad$ and then restrict these degrees of freedom to finite sub-regions $\Rh$ of $\scrip$ to obtain $\psRh$. In section \ref{s3.2} we show that the BMS vector fields $\xi^a$ induce infinitesimal canonical transformations, providing us with Hamiltonians $H_\xi [\Rh]$ on $\psRh$ that represent fluxes $F_\xi [\Rh]$ of the BMS momenta across $\Rh$. Following Refs.~\cite{aams,Dray1984,Dray1985}, in section \ref{s3.3} we obtain the expressions of BMS charges by `integrating' the fluxes. The passage from fluxes to charges requires one to step out of the radiative phase space and use full Einstein equations and Bianchi identities at $\scrip$ that relate fields carrying `Coulombic' and `radiative' information at $\scrip$. However, in contrast to other approaches to charges and fluxes \cite{PhysRev.150.1039,Geroch:1981ut,wz,cfp}, the entire analysis can be carried out at $\scrip$ without the need of extending symmetries or physical fields to the space-time interior.

\subsection{Radiative phase spaces at $\scrip$} 
\label{s3.1}
 
To discuss charges and fluxes associated with symmetries, we have to work with the phase space consisting of all space-times that admit $\scrip$ as a boundary in their conformal completion.  Let us then begin with the phase space $\ps$ of all vacuum solutions $(M, g_{ab})$ of Einstein's equations that are asymptotically flat at null infinity (in the sense of Definition 2 of \cite{aass1}), and work with their conformal completions $(\hM,\,\hg_{ab})$ in which $\scrip$ is divergence-free. Then $\scrip$ is equipped with a kinematical structure that it is shared by all space-times in $\ps$: Pairs $(\hq_{ab},\, \hn^a)$ where any two are conformally related via $(\hq^\prime_{ab} = \mu^2 \hq_{ab},\, \hn^{\prime\,a} = \mu^{-2} \hn^a)$ with $\mu$ satisfying $\Lie_{\hn} \mu =0$. The role of this kinematic structure is analogous to that played by the Minkowski metric in section \ref{s2}.

As discussed in \cite{aass1}, the dynamical information in the gravitational field at $\scrip$  is encoded in the intrinsic connection $\hD$, induced by the space-time connection $\h\nabla$. $\hD$ is compatible with the kinematic structure --i.e., satisfies $\hD_a \hq_{bc} \= 0$ and $\hD_a \hn^b \=0$-- but varies from one space-time to another. It captures the radiative information in the physical metric $g_{ab}$ that is locally registered at $\scrip$ and is thus analogous to the radiation field $\h\phi$ of section \ref{s2}. The phase space $\psrad$ adapted to $\scrip$ consists of these degrees of freedom.

More precisely, we have the following structure \cite{aa-radiativemodes,ashtekar1987asymptotic}. The curvature of $\hD$ on $\scrip$ encodes the Bondi news $\h{N}_{ab}$ and the part ${}^\star\!\h{K}^{ab} := \lim_{{\scrip}}\, (\Omega^{-1}\, {}^\star\h{C}^{acbd} \,\hn_c\hn_d)$ of the asymptotic Weyl curvature of $\h{g}_{ab}$. The five components of ${}^\star\!\h{K}^{ab}$ encode the radiative modes corresponding to the Newman-Penrose components $\Psi_4^\circ,\, \Psi_3^\circ,\, {\rm Im}\Psi_2^\circ$, all of which vanish in any stationary space-time. Because every $\hD$ annihilates both $\hq_{ab}$ and $\hn^a$, any 2 of our derivative operators $\hD$ and ${\hD}{}^\prime$ on $\scrip$ are related by 
\be \label{diff} (\hD_a - {\hD}{}^\prime_a)\hat{f}_b = n^c \hat{f}_c\, \h\Sigma_{ab}  \qquad   \hbox{\rm for arbitrary 1-forms $\hat{f}_b$ on $\scrip$}\, , \ee
and some  symmetric tensor field $\h\Sigma_{ab}$ that is transverse to $\hn^a$,\, i.e., satisfies $\h\Sigma_{ab} \hn^a \=0$. This property brings out an important fact: the action of all derivative operators is the same on `horizontal' 1-forms $\hat{h}_a$ satisfying $\hat{h}_a \hn^a \=0$. Therefore,\, (as discussed in \cite{aass1})\, $\hD$ is characterized by its action on any one 1-form $\h\ell_a$ on $\scrip$ satisfying $\hn^a\, \h\ell_a = -1$. This fact plays an important role in phase space considerations.

But there is gauge freedom in $\hD$: If we change the conformal factor by $\Omega \to \Omega^\prime = \mu\,\Omega \equiv  (1+ \Omega f)\Omega$ in a neighborhood of $\scrip$, then clearly $\h{q}_{ab}$ and $\hn^a$ do not change, but $\hD \to \hD^\prime$ with $\h\Sigma_{ab} \= f \hq_{ab}$. Thus the `pure-trace part' of $\h\Sigma_{ab}$ is gauge and we are led to consider two connection as equivalent if they differ just by this gauge freedom:
\be \hD\, , {\hD}{}^\prime\, \in \{\hD\} \qquad {\rm iff} \quad (\hD_a -{\hD}{}^\prime_a )\hat{f}_b \,\propto\, (n^c \hat{f}_c) \hq_{ab}\, . \ee

Therefore, the difference between any two equivalence classes is characterized by the trace-free part $\h{\mathfrak{f}}_{ab} := \h\Sigma_{ab} - \f{1}{2} \h{q}^{cd} \h\Sigma_{cd}\, \hq_{ab}$, where $\h{q}^{cd}$ is any inverse of $\hq_{ab}$. Note that this transverse traceless $\h{\mathfrak{f}}_{ab}$ has precisely two components; they represent the two radiative modes of the gravitational field. It is remarkable that they can be isolated locally in a simple way in spite of the non-linearities of full general relativity; this is why the $\scri$-framework is so well-adapted to the study of radiation fields. Note however that for this very reason the connections $\{\hD\}$ have no knowledge of the Coulombic aspects of the gravitational field that are encoded, e.g., in\, ${\rm Re} \Psi_2^\circ$ that enters the expression of the Bondi-Sachs 4-momentum and $\Psi_1^\circ$ that enters the expression of the angular momentum charge. In terms of the discussion of section \ref{s2} of \cite{aass1}, the degrees of freedom captured by $\{\hD\}$ are `freely specifiable' fields on 3-dimensional $\scrip$. The Coulombic information can be specified on a 2-sphere (say in the distant past), and is then determined everywhere on $\scrip$ by the radiative modes. It does \emph{not} have `dynamical content' that freely specifiable fields have; it constitutes the `corner data'.

The phase space that captures the degrees of freedom in $g_{ab}$ that are relevant to $\scrip$ is thus the space $\psrad$ of connections $\{\hD\}$ at $\scrip$ (subject to appropriate regularity conditions specified in \cite{aams}). Therefore it has the structure of an affine space. For intermediate calculations, it is convenient to endow it a vector space-structure by choosing an origin. Fortunately, there is a natural class  of connections that can serve as origins: $\{\vac\}$ that have trivial curvature, i.e., for which ${}^\star\!\h{K}^{ab} =0$ (which implies that $\h{N}_{ab}$ also vanishes \cite{abk}). These points $\{\vac\} \in \psrad$ are referred to as \emph{`classical vacua'}. Each classical vacuum $\{\vac\}$ is left invariant by the translation subgroup $\T$ of the BMS group $\B$, and the quotient $\S/\T$ of the supertranslation group by its translation subgroup acts freely and transitively on the space of classical vacua; thus there as `as many' classical vacua $\vac$ as there are supertranslations modulo translations. Let us fix a $\{\vac\}$ and choose it as the origin in $\psrad$. Then, the difference between an arbitrary element $\{\hD\}$ and $\{\vac\}$ is completely characterized by a transverse, traceless, symmetric tensor field $\h\gamma_{ab}$ \cite{aa-radiativemodes}: 

\be \label{gamma} \h\gamma_{ab} = \big(\{ \hD \}_a - \{ \vac \}_a\big)\h\ell_b = {\rm TF}(\hD_a - \vac_a)\h\ell_b\, . \ee
Here TF stands for ``trace-free part of", $\hD$ is any element of $\{ \hD \}$ and $\vac$ of $\{ \vac \}$, and, as before, $\h\ell_b$ is any 1-form on $\scrip$ satisfying $\hn^b \h\ell_b = -1$. The fields $\h\gamma_{ab}$ serve as convenient `coordinates' in the infinite dimensional affine space $\psrad$. We will refer to $\h\gamma_{ab}$ as \emph{relative shear} because it represents the difference between (generalized) shears assigned to $\h\ell_a$ by the phase space variable $\{ \hD \}$ and the vacuum $\{ \vac \}$. (`Generalized' shears because $\h\ell_a$ does not have to be orthogonal to a cross-section). Note that the relative shear is independent of the choice of the 1-form $\h\ell_a$ on $\scrip$ satisfying $\hn^a \h\ell_a =-1$. It is easy to verify that $\h\gamma_{ab}$ has conformal weight $1$\, --i.e., under $\hq_{ab} \to \mu^2 \hq_{ab}$ we have $\h\gamma_{ab} \to \mu\hat \gamma_{ab}$-- \,which will ensure that the final results are conformally invariant. Our results will be insensitive to the choice of $\{\vac\}$ that serves as the `origin'. Finally, one can easily  show that the relative shear is a natural potential for the News tensor defined by $\{\hD\}$
\cite{aa-radiativemodes,aams}: 
\be \label{newspotential} \h{N}_{ab}\, = 2\,\Lie_{\hn} \h\gamma_{ab} .\ee
The dependence to the `origin' $\{ \vac \}$ in the definition of $\h\gamma_{ab}$ drops out when we take its $\Lie_{\hn}$. \medskip

\emph{Remarks:} 

1. There is a useful analogy with Minkowski space, which is also an affine space: choice of $\{\vac\}$ is the analog of the choice of an origin in Minkowski space that is generally made to introduce Cartesian coordinates $x^{\mathfrak{a}}$, and $\h\gamma_{ab}$ are the analogs of these Cartesian coordinates. Under the change of the origin, the Cartesian coordinates $x^{\mathfrak{a}}$ in Minkowski space undergo a constant shift $x^{\mathfrak{a}} \to x^{\prime \, \mathfrak{a}} = x^{\mathfrak{a}} + c^{\mathfrak{a}}_{\circ}$. Similarly, under $\vac \to {\vac}{}^\prime$, the coordinate $\h\gamma_{ab}$ undergoes a constant shift, $\h\gamma_{ab} \, \to\, \h\gamma_{ab} + \h{c}_{ab}^\circ$. The tensor field $\h{c}_{ab}^\circ$ on $\scrip$ has the form $\h{c}_{ab}^\circ =\, {\rm TF}\,\big(\hD_a \hD_b\, \super +  \f{\super}{2}\,\h\rho_{ab}\big)$ where $\super$ denotes the supertranslation relating the two vacua $\{ \vac \}$ and $\{ \vac{}^\prime \}$, and $\h\rho_{ab}$ is the Geroch tensor field constructed from the kinematical structure \cite{gerochrev}.\, $\h{c}_{ab}^\circ$ satisfies $\Lie_n \h{c}_{ab}^\circ =0$; it has no dynamical content. It vanishes if and only of $\super\hn^a$ is a translation.   \medskip

2. Although the relative shear $\hat\gamma_{ab}$ is symmetric and trace-free, conceptually it is quite different from the more familiar shear $\h\sigma_{ab}$ of cross-sections of $\scrip$.\, The former does not refer to any cross-section but requires a choice of classical vacuum $\{\vac \}$, while the latter does not refer to any vacuum $\{\vac \}$ but requires a choice of cross section. Choosing an $\h\ell_a$ orthogonal to a cross-section, one can relate the two using the identity
\be\label{Dltoshear}
\h{D}_a \h\ell_b = \h{q}_a{}^c\, \h{q}_b{}^d \,\hD_c \h\ell_d\, -\, \h\ell_a\h\tau_b\, \equiv \,
\h\sigma_{ab}\,+ \,\f{1}{2}\h\theta\, \hq_{ab}\, -\,\h\ell_a\h\tau_b\, ,
\ee
where $\h{q}_{a}{}^{c}$ is projector on the tangent space of the chosen cross-section, $\h\tau_a:=\Lie_{\hn} \h\ell_a$ and $\h\theta$ is the expansion of $\h\ell_a$. Then the difference $(\{ D_a\} -\{\vac_a\}) \h\ell_b$  defining $\h\gamma_{ab}$ corresponds to subtracting from $\h\sigma_{ab}$ shear w.r.t. a vacuum connection $\{\vac\}$. 

3. Definition of the relative shear  brings out the fact that $\psrad$ can be regarded as a (trivial) fibre bundle in which the base space is spanned by `non-dynamical' classical vacua $\{ \vac \}$ and the fibers are coordinatized by the curvature ${}^\star\!\h{K}^{ab}$ of $\{\hD\}$.  Given a conformal completion, a change $\{\vac_a\}\,\to\, \{ \vac{}^\prime \}$ in the choice of the `origin' can be regarded as a `Goldstone mode', \`a la \cite{Strominger:2013jfa}.
$\h\gamma_{ab}$ is also related to the `covariant shear' of \cite{Compere:2018ylh}. \\

The symplectic structure $\bfomega$ on $\ps$ \cite{cw,abr,wz} induces a natural $\bfomegarad$ on $\psrad$ \cite{aaam-cmp}. Given any point $\{\hD\} \in \psrad$,  tangent vectors $\delta$ at $\{\hD\}$ are represented by fields $\delta\h\gamma_{ab}$ on $\scrip$ that are by definition symmetric, transverse and trace-free. The action of $\bfomegarad$ is given by \cite{aaam-cmp,aams} 
\be \label{symp4} \bfomegarad\!\mid_{{}_{\{\hD\}}} (\delta_1, \delta_2) = \frac{1}{8\pi G}\,\int_{\scrip} \Big[(\delta_1\, \h\gamma_{ab})\, (\Lie_{\hn}\, \delta_2\, \h\gamma_{cd})\, - \, (\delta_2\, \h\gamma_{ab}) (\Lie_{\hn}\, \delta_1 \,\h\gamma_{cd})\, \Big]\, \hq^{ac}\, \hq^{bd}\,\, \rmd^3\! \scrip \ee
where $\hq^{ab}$ is \emph{any} inverse of $\hq_{ab}$. Because $\delta\h\gamma_{ab},\, \hn^a,\, \h{q}^{ab}$ and the volume element $\rmd^3\! \scrip$ have conformal weights $1,\, -1,\, -2$ and $3$ respectively, it follows that the integral on the right side is conformally invariant. It is also invariant under the change of origin in the affine space $\psrad$ because each tangent vector $\delta\h\gamma_{ab}$ is itself invariant under this displacement. 

We can now introduce the phase space $\psRh$ associated with open regions $\Rh$ in $\scrip$, bounded by two 2-sphere cross-sections. As before, one begins by restricting the phase space variable $\{\hD\}$ to $\Rh$ and introducing a suitable norm on it to specify topology:
\be  \label{norm2}||\h\gamma||^2_{{}_{\Rh}}\, := \, \int_{\Rh} \big[ |\h{n}^a \hD_a \h\gamma_{bc}|^2 + \hq^{ab}\hq^{cd}\hq^{mn}\, \hD_a \h\gamma_{cm}\, \hD_b \h\gamma_{dn} + \f{1}{l^2} |\h\gamma_{ab}|^2\big]\, \rmd^3\! \scrip \ee
where the conventions are the same as in Eq.~(\ref{norm1}). 
Again, we will only use the topology induced on $\psRh$ by this norm, and the topology is insensitive to the choices of auxiliary structures used Eq.~(\ref{norm2}). The symplectic structure $\bfomegaRh$ on $\psRh$ is obtained by restricting the integral on the right side of (\ref{symp4}) to $\Rh$. It is again weakly non-degenerate and continuous on $\psRh$. 

Note that there are two interesting contrasts with the phase space $\ps$. First, while $\ps$ is a genuinely non-linear space, $\psrad$ and $\psRh$ have a natural affine space structure. Second, while $\bfomega$  on $\ps$ has degenerate directions that correspond to pure-gauge linearized fields\, --it is only a pre-symplectic structure--\, $\bfomegarad$ and $\bfomegaRh$ are both non-degenerate; there is no gauge freedom because $\{\hD\}$ contains only the two physical, radiative degrees of freedom of full general relativity. These key simplifications will be exploited in out approach to obtaining fluxes and charges.

\subsection{BMS Fluxes}
\label{s3.2}

Next, let us consider the action of the BMS vector fields $\xi^a$ on elements $\{\hD\}$ of $\psRh$. For simplicity of presentation let us first consider the BMS symmetries which preserve the given conformal frame, i.e., satisfy $\Lie_{\xi}\hq_{ab} \=0$ and $\Lie_{\xi} \hn^a \=0$ \,(these are linear combinations of supertranslations and rotational Killing fields of $\h{q}_{ab}$),\, and discuss general BMS symmetries\, (i.e., those that also contain a boost)\, at the end. 

The vector $\delta_\xi$ at a point $\{ \hD\}$ of $\psrad$ represents the infinitesimal change in that $\{\hD\}$ under the diffeomorphism generated by $\xi^a$. Recall that $\{\hD\}$ is completely determined by its action $\{\hD_a\ell_b\}$ on a 1-form  $\h\ell_b$ satisfying $\h\ell_b \hn^b =-1$. Therefore, the change $\delta_\xi\{\hD\}$ in $\{\hD\}$ is faithfully encoded in ${\rm TF} (\Lie_\xi \hD_a - \hD_a \Lie_\xi)\h\ell_b$. Under this infinitesimal motion on $\psrad$, the coordinate labels $\h\gamma_{ab}$ shift by
\be  \label{deltaxi}  \delta_\xi\h\gamma_{ab}  = {\rm TF} (\Lie_\xi \hD_a - \hD_a \Lie_\xi)\h\ell_b\, , \ee
providing us  the `components' of the vector field $\delta_\xi$ at the point $\{\hD\}\in \psrad$. (Since \smash{$(\Lie_\xi \hD_a - \hD_a \Lie_\xi)\h\ell_b$} is automatically symmetric and transverse to $\hn^a$, the operation of removing trace is well-defined.) Note that the origin $\vac$ is kept fixed since it not the dynamical variable:\, we are computing the `components' of $\delta_\xi $ in fixed `coordinates' $\hat\gamma_{ab}$ on $\psrad$. 

Next, we note that 
\begin{align} \label{difference} \Lie_\xi \h\gamma_{ab} &= {\rm TF}\, \Lie_\xi [(\hD_a- \vac_a)\h\ell_b] 
= {\rm TF}\, [(\Lie_\xi \hD_a- \hD_a \Lie_\xi)\h\ell_b - (\Lie_\xi\vac_a- \vac_a \Lie_\xi)\h\ell_b] 
\nn\\&= \delta_\xi \h\gamma_{ab} - {\rm TF}\,(\Lie_\xi \vac_a - \vac_a \Lie_\xi)\,\h\ell_b \, .\end{align}
In the first step we have set $\hD_a (\Lie_\xi \h\ell_b) = \vac_a (\Lie_\xi \h\ell_b)$ using the fact that all derivative operators have the same action on $\Lie_\xi \h\ell_a$ because $\hn^a (\Lie_\xi \h\ell_a) =0$.
We conclude that \smash{$\delta_\xi \h\gamma_{ab} \not= \Lie_\xi \h\gamma_{ab}$.} 
\footnote{Such mismatches are sometimes referred to as `anomalies' in the literature; see e.g. \cite{Hopfmuller:2018fni,Chandrasekaran:2020wwn,Freidel:2021cjp,Odak:2022ndm}. The inequality $\delta_\xi \h\gamma_{ab} \not= \Lie_\xi \h\gamma_{ab}$ simply encodes the fact that $\h\gamma_{ab}$ is a functional of both dynamical and background fields and $\delta_\xi$ acts only on dynamical fields while $\Lie_\xi$ acts on all fields.}
Next, the action of the BMS group maps a connection $\hD$ to another connection $\hD^\prime$ and the curvature ${}^\star\!\h{K}^{ab}$ of $\hD$ is therefore sent to that of $\hD^\prime$. Since vacuum connections are characterized by vanishing of ${}^\star\!\h{K}^{ab} =0$, it follows that $\vac$ is mapped to another vacuum connection ${\vac}{}^\prime$. 
(In terms of the analogy discussed in Remark 1 at the end of section \ref{s3.1}, this is the analog of a constant shift of origin in Minkowski space-time.) Since all vacua are related by a BMS supertranslation $\super\hn^a$ it follows that  ${\rm TF}\,(\Lie_\xi \vac_a - \vac_a \Lie_\xi)\h\ell_b = {\rm TF}\, (\vac_a \vac_b \super  + \f{1}{2}\,\super \h\rho_{ab} )$ for some $\super$ satisfying $\Lie_{\hn} \super =0$. Next, because $\vac_b \super$ is a horizontal 1-form, $\Lie_{\hn}\, (\vac_a \vac_b \super) = \vac_a (\Lie_{\hn} \vac_b \super) =0$ and, by its definition, the Geroch tensor satisfies $\Lie_{\hn} \h\rho_{ab} =0$.} Therefore $\Lie_{\hn} \big((\Lie_\xi \vac_a - \vac_a \Lie_\xi)\h\ell_b\big) =0$. Hence, although $ \delta_\xi \h\gamma_{ab} \not= \Lie_\xi \h\gamma_{ab}$, we do have 
\be \label{Liedelta}\Lie_{\hn} \big( \delta_\xi \h\gamma_{ab} - \Lie_\xi \h\gamma_{ab}\big) = 0. \ee
We also have that for \emph{any} vector field $\delta$ on $\psRh$
\be \label{deltadelta} 
\delta\, \big((\delta_\xi -\Lie_\xi)\h\gamma_{ab}\big)= 0,
\ee
 because $\vac_a \vac_b \super  + \f{1}{2}\,\super \h\rho_{ab}$ does not depend on the dynamical variable $\hD$ on $\psRh$.
(More precisely, the Lie-derivative of  functions $\int_{\Rh} f^{ab} (\delta_\xi -\Lie_\xi)\h\gamma_{ab}$ along any vector field $\delta$ on $\psRh$ vanishes for all (field independent) test fields $f^{ab}$ on $\Rh$.) These properties will be useful below.

To determine if the infinitesimal motion $\delta_\xi \h\gamma_{ab}$ on $\psRh$ is Hamiltonian, we need to evaluate $\bfomegaRh (\delta_\xi, \delta)$ and check if there exists a function $H_\xi$ on $\psRh$ such that $\bfomegaRh (\delta_\xi, \delta) = \delta H_\xi$ for all vector fields $\delta$. As in section \ref{s2}, to carry out this check it suffices to let $\delta$ be an arbitrary \emph{constant} vector field $\delta_\circ$ on the affine space $\psRh$ because constant vector fields span the tangent space at every point of $\psRh$.\, A vector field $\delta_\circ$ on $\psR$ is  said to be constant if:\\ 
(i) It shifts every phase space point $\{\hD\}$ by the same amount:  $\delta_\circ \big(\{ \hD \}_a \h\ell_b\big) = \h\gamma_{ab}^\circ$, a fixed tensor field $\h\gamma_{ab}^\circ$ on $\Rh$;\, or, equivalently,\\
(ii) The action of $\delta_\circ$ on the phase space `coordinates' is given by $\delta_\circ \h\gamma_{ab} = \h\gamma_{ab}^\circ$, so that the infinitesimal constant shift is $\h\gamma_{ab} \to \h\gamma_{ab} + \epsilon \h\gamma_{ab}^\circ$ for all $\h\gamma_{ab}$.\\
Here the tensor field $\h\gamma_{ab}^\circ$ on $\psRh$ is subject only to the condition that the norm $||\h\gamma_{ab}^\circ||$ given by (\ref{norm2}) is finite.

With these preliminaries out of the way, let us return to the question of whether the vector field $\delta_\xi$ is Hamiltonian. As in the case of a scalar field $\phi$ of section \ref{s2}, the vector field $\delta_\xi$ is well defined on the dense subspace $\DRh$ of $\psRh$ on which\, $ \delta_\xi \h\gamma_{ab} = {\rm TF}\,(\Lie_\xi \hD_a - \hD_a \Lie_\xi)\h\ell_b$\, has a finite norm (\ref{norm2}),\, and  at any point $\{\hD\}$ in $\DRh$ we have:
\be \bfomegaRh\!\mid_{{}_{\{\hD\}}} (\delta_\xi, \delta_\circ) = \frac{1}{8\pi G} \, \int_{\Rh} \Big[(\delta_\xi \h\gamma_{ab}) (\Lie_{\hn}\, (\delta_\circ\h\gamma_{cd})) \,-\, (\delta_\circ \h\gamma_{ab}) (\Lie_{\hn} (\Lie_\xi \h\gamma_{cd}))\big]\, \h{q}^{ac} \h{q}^{bd}\, \rmd^3\! \scrip\, , \ee 
where we have used (\ref{Liedelta}) in the second term on the right side. Next we use the property $\Lie_\xi \hn^a =0$ of the BMS vector field $\xi^a$ under consideration and integrate this term by parts to obtain
\ba \label{HamVF1}
\bfomegaRh\!\mid_{{}_{\{\hD\}}} (\delta_\xi, \delta_\circ) &=& \frac{1}{8\pi G} \, \int_{\Rh} \Big[(\delta_\xi \h\gamma_{ab}) (\Lie_{\hn}\, (\delta_\circ \h\gamma_{cd})) \,+\, (\Lie_{\xi}\,(\delta_\circ \h\gamma_{ab})) (\Lie_{\hn} \h\gamma_{cd})\big]\, \h{q}^{ac} \h{q}^{bd}\,  \rmd^3\! \scrip\, \nonumber\\ 
 &-& \frac{1}{8\pi G} \, \oint_{\partial\Rh} \big[(\delta_\circ \h\gamma_{ab})(\Lie_{\hn}\h\gamma_{cd})\, \h{q}^{ac} \h{q}^{bd}\, \xi^m\big]\, \rmd^2\h{S}_m \, . \ea
As in section \ref{s2}, the surface term vanishes if $\xi^a$ happens to be tangential to $\partial\Rh$. If not, as before  we further restrict $\DRh$ to the dense subspace 
$\DRhp$
 of $\psRh$ on which $\Lie_{\hn}\h\gamma_{cd}$ vanishes on $\partial\Rh$ (i.e., the news tensor $\h{N}_{ab}$ vanishes there; see Eq.~(\ref{newspotential})). On this
 $\DRhp$ only the volume term on the right hand side of (\ref{HamVF1}) survives. \smallskip

Next we note two properties of constant vector fields $\delta_\circ$. Using the fact that the action of $\delta_\circ$ on phase space coordinates $\h\gamma_{ab}$ commutes with the action of Lie derivatives w.r.t. vector fields that do not depend on phase space variables, we have \\
(i) $\Lie_{\hn} (\delta_\circ\, \h\gamma_{cd}) = \delta_\circ (\Lie_{\hn}\h\gamma_{cd})$;\, and,\\
(ii) $\Lie_\xi (\delta_\circ\, \h\gamma_{ab}) = \delta_\circ\, (\Lie_\xi \h\gamma_{ab}) = \delta_\circ\, (\delta_\xi \h \gamma_{ab})$, where in the last step we have used (\ref{deltadelta}).
Therefore on the dense subspace $\DRhp$ the right side of (\ref{HamVF1}) can be rewritten as 
\ba \bfomegaRh\!\mid_{{}_{\{\hD\}}} (\delta_\xi, \delta_\circ) &=& \frac{1}{8\pi G} \int_{\Rh} \Big[(\delta_\xi \h\gamma_{ab}) (\delta_\circ (\Lie_{\hn}  \h\gamma_{cd})) \,+\, (\delta_\circ(\delta_{\xi} \h\gamma_{ab})) (\Lie_{\hn} \h\gamma_{cd})\big]\, \h{q}^{ac} \h{q}^{bd}\,  \rmd^3\! \scrip\, \nonumber\\ 
&=& \frac{1}{8\pi G}\,\,\delta_\circ \int_{\Rh} (\delta_\xi \h\gamma_{ab})(\Lie_{\hn}\h\gamma_{cd})\, \hq^{ac}\hq^{bd}\,\, \rmd^3\! \scrip\,\ea
Therefore we conclude that on the dense subspace $\mathcal{D}^\prime_{\Rh}$ of $\psRh$, the Hamiltonian $H_\xi$ is given by
\be H_\xi (\{\hD\}) = \f{1}{8\pi G}\, \int_{\Rh} (\delta_\xi \h\gamma_{ab})(\Lie_{\hn}\h\gamma_{cd})\, \hq^{ac}\hq^{bd}\,\, \rmd^3\! \scrip\, ,\ee
where we have eliminated the freedom to add a constant by requiring that $H_\xi$ should vanish on any classical vacuum. (Recall that $\Lie_{\hn}\h\gamma_{cd} = 2 \h{N}_{ab}$ vanishes if $\{\hD\}$ is a classical vacuum.) 

By inspection, the expression of the function $H_\xi$ on the dense space $\DRhp$ is continuous and therefore it admits a continuous extension to all of $\psRh$. This is the Hamiltonian that generates the infinitesimal canonical transformation $\delta_\xi$. As in section \ref{s3}, the Hamiltonian is defined on the full phase space $\psRh$ but the Hamiltonian vector field it generates is defined only on a dense subspace. This is a common occurrence on infinite dimensional phase spaces. Also, as in section \ref{s3}, \emph{if} $\xi^a$ is tangential to the boundary $\partial\Rh$, the 1-parameter family of diffeomorphisms generated by $\xi^a$ on $\Rh$ induces a 1-parameter family of canonical transformations on full $\psRh$. If not, we only have an infinitesimal canonical transformation $\delta_\xi$, again a common occurrence in infinite dimensions. 

Finally, using the expression (\ref{deltaxi}) of $\delta_\xi\h \gamma_{ab}$, and the relation (\ref{newspotential}) between $\gamma_{ab}$ and the News tensor $\h{N}_{ab}$ of $\{\hD\}$, we can rewrite the Hamiltonian directly in terms of the phase space variable $\{\hD\}$, without reference to the phase-space coordinate $\h\gamma_{ab}$. We have:
\be \label{H1bis} H_\xi (\{\hD\}) = \f{1}{16\pi G}\,\int_{\Rh} \big[(\Lie_{\xi} \hD_a - \hD_a \Lie_{\xi})\h\ell_b \big]\,\h{N}_{cd}\, \hq^{ac}\hq^{bd}\,\, \rmd^3\! \scrip\, , \ee
where we could drop the qualifier ${\rm TF}$ in front of the term in the square bracket because the News tensor is already trace-free. Note that the term $(\Lie_{\xi} \hD_a - \hD_a \Lie_{\xi})\h\ell_b$ is precisely the action on $\ell_b$ of the infinitesimal change in $\hD$ under the action of the diffeomorphism generated by the BMS vector field $\xi^a$. 

So far we restricted ourselves to BMS vector fields $\xi^a$ that preserve the conformal frame $(\h{q}_{ab},\, \hn^a)$. These include supertranslations $\xi^a = \super \hn^a$. For these vector fields, the expression of the Hamiltonian can be simplified and one has \cite{aams} 
\be \label{H2}  H_{\super} (\{\hD\}) = \f{1}{32\pi G} \,\,\int_{\Rh} \big[\super\,\h{N}_{ab}\, +\, 2\hD_a \hD_b\, \super + \super\,\h\rho_{ab}\big]\, \h{N}_{cd}\,\hq^{ac}\hq^{bd}\,\, \rmd^3\! \scrip\, .\ee 
The term quadratic in their news tensor is often called the `hard contribution' to the supermomentum flux, and the term linear in news, the `soft contribution'. For BMS translations $\tran\, n^a$, the coefficient $\tran$ satisfies ${\rm TF} \big(2\hD_a \hD_b \tran + \tran\,\h\rho_{ab}\big) =0$, whence the `soft contribution' vanishes and the flux of the Bondi-Sachs 4-momentum across $\Rh$ is given by: 
\be\label{H3}   H_{\tran} (\{\hD\}) = \f{1}{32\pi G} \,\,\int_{\Rh} \tran\,\h{N}_{ab}\, \h{N}_{cd}\,\,\hq^{ac}\hq^{bd}\,\,   \rmd^3\! \scrip\, \ee 

Finally, let us consider a general BMS vector field $\xi^a$, $\Lie_{\xi} \h{q}_{ab} = 2\beta\, \hq_{ab}$ and $\Lie_\xi \hn^a = - \beta\, \hn^a$. Recall that $\h\gamma_{ab}$ has conformal weight $1$ and, as we noted above, if $\beta=0$, then\, $\delta_\xi \h\gamma_{ab} = {\rm TF} (\Lie_\xi \hD_a - \hD_a \Lie_\xi)\h\ell_b$\, again has conformal weight $1$. If $\beta\not=0$, then ${\rm TF} (\Lie_\xi \hD_a - \hD_a \Lie_\xi)\h\ell_b$ no longer has conformal weight $1$ because the diffeomorphism generated by $\xi^a$ changes the conformal frame. Therefore, as in the case of the scalar field $\h\phi$ of section \ref{s3.2}, the expression of $\delta_{\xi}$ acquires an extra term:
\be \delta_{\xi} \h\gamma_{ab} = {\rm TF} \big[(\Lie_\xi \hD_a - \hD_a \Lie_\xi)\h\ell_b \, +\, 2 \h\ell_{(a} \hD_{b)}\beta \big]\, . \ee
The extra term ensures that $\delta_{\xi} \h\gamma_{ab}$ is again of conformal weight $1$ (and symmetric, transverse and traceless), and therefore qualifies as a tangent vector to $\psRh$.%
\footnote{Thus, the situation is completely analogous to that we encountered in section \ref{s2.2}. The additional term can again be systematically arrived at by examining the transformation property of the phase space variable under conformal rescalings. On $\psRh$ now under consideration, under a (finite) conformal rescaling $\hq_{ab} \to \hq^\prime_{ab} = \mu^2 \hq_{ab}$ and $\hn^a \to \hn^{\prime\, a} = \mu^{-1} \hn^a$ we have $\{ {\hD}{}^\prime_a \} \hat{f}_b = \{ {\hD}_a \} \hat{f}_b - 2 \mu^{-1} (\hD_{(a} \mu)\, \hat{f}_{b)}$.} 
With this expression of $\delta_\xi$ at hand, one can repeat, step by step, the above procedure to arrive at Hamiltonians $H_\xi$, for general BMS vector fields $\xi^a$. However, now the intermediate expressions are much
longer because the diffeomorphisms generated by $\xi^a$ change the conformal frame. In particular, several terms that vanished for BMS vector fields whose action preserves the conformal frame no longer do. For example, now $(\hD_a - \vac_a)\, \Lie_\xi\h\ell_b \not=0$  because $\Lie_\xi\h\ell_b$ is no longer a horizontal 1-form (i.e., $\hn^b\,\Lie_\xi \h\ell_b = \Lie_\xi \beta \not=0$).  However, the final expression of the Hamiltonian $H_\xi$ is rather simple and transparent:
\be \label{H4} H_\xi (\{\hD\}) = \f{1}{16\pi G}\,\,\int_{\Rh} \big[(\Lie_{\xi} \hD_a - \hD_a \Lie_{\xi})\h\ell_b + 2 \h\ell_{(a} \hD_{b)} \beta \big]\,\h{N}_{cd}\, \hq^{ac}\hq^{bd}\,\, \rmd^3\! \scrip \, .\ee

As before, we made three choices to obtain this expression of $H_\xi (\{\hD\})$: a conformal frame $(\hq_{ab}, \hn^a)$, a 1-form $\h\ell_a$ satisfying $\h\ell_a \hn^a =-1$, and an inverse metric $\hq^{ab}$. One can directly check that the integral on the right side of (\ref{H4}) is independent of all these choices. As usual there is freedom to add a constant to  $H_\xi$ and, as in section \ref{s3.2}, we have eliminated it by asking that $H_\xi$ should vanish  at vacuum configuration $\{\vac\} \in \psRh$. \emph{Motivated by our findings for the scalar field $\h\phi$ in section \ref{s3.2}, we are led to interpret $H_\xi$ as the flux  $F_\xi [\hR]$ across $\Rh$ of the component of the BMS momentum defined by $\xi^a$.} \medskip

\emph{Remarks:}

1. In the literature, the expression of the Hamiltonian $H_\xi$ is often written using shear $\h\sigma_{ab}$ (in place of the derivative operator $\hD$) and equating $\h{N}_{ab}$ with $2\, {\rm TF}\, \Lie_{\hn} \h\sigma_{ab}$, where ${\rm TF}$ stands for `trace-free part of', trace being taken using the intrinsic 2-metric on  cross-sections. This equality does not hold without further qualifications since  $\h{N}_{ab}$ is defined intrinsically on $\scrip$ \cite{gerochrev}, while $\h\sigma_{ab}$ generally refers to a foliation. Given a foliation to which $\h\ell_b$ is orthogonal, the relation between $\h\sigma_{ab}$ and $\h{N}_{ab}$ can be derived starting from \eqref{Dltoshear} and properties of the Schouten tensor of $\h{g}_{ab}$. We have:\,\,  
\be \h{N}_{ab} =  2 \Lie_{\hn} \h\gamma_{ab} = {\rm TF} \big[2\Lie_{\hn} \h\sigma_{ab} - 2(\h{D}_{( a}+\h\tau_{(a})\h\tau_{b)} - 2 \h\ell_{(a}\, \Lie_{\h{n}}\, \h\tau_{b)} - \h\rho_{ab}\big].\ee
This general expression can be simplified by making special choices of the conformal frame $(\h{q}_{ab}, \hn^a)$ on $\scrip$ and the 1-form $\h\ell_a$ satisfying $\hn^a\h\ell_a =-1$. In particular, if $\h{q}_{ab}$ is chosen to be a unit, round 2-sphere metric, and $\h\ell_a$ is chosen so that  $\Lie_{\h{n}} \h\ell_b =0$, then $\hat\tau_a =0$ and $\h\rho_{ab} = \hq_{ab}$, then we would have $\h{N}_{ab} = 2 \Lie_{\hn} \h\sigma_{ab}$. In a general conformal frame ($\hq_{ab}^\prime = \omega^2 \hq_{ab},\, \hn^{\prime}{}^a = \omega^{-1} \hn^a$) we would also obtain $\h{N}_{ab} = 2 \Lie_{\hn} \h\sigma_{ab}$,\, \emph{provided} we restrict $\h\ell^{\prime}_a$ to be tied to the $\h\ell_a$ used in the Bondi frame via $\h\ell^{\prime}_a = \omega \h\ell_a$. However, these simplifications do not occur for generic choices. For example, if in a non-Bondi frame $\hq^\prime_{ab}$, one were to choose $\h\ell^{\prime}_a$  satisfying ${\hat\tau}^\prime_a =0$, we would have $\h{N}_{ab} = {\rm TF} \big[2\Lie_{\hn}\h\sigma_{ab}^\prime -\h\rho^\prime_{ab}]$ and $\rho^\prime_{ab}$ would not be pure trace.   

2. The fluxes associated with every BMS vector field $\xi^a$ vanish identically on \emph{any} region $\Rh$ if the Bondi news $\h{N}_{ab}$ vanishes there; the 2-spheres bounding $\Rh$ are not tied to any specific (e.g., Bondi-type) foliation. While this is clearly a desired physical property, it is not shared by all flux expressions available in the literature. In particular, the flux resulting from `linkages' does not have this property \cite{Geroch:1981ut,10.1063/1.525283}. (Note also that if, contrary to Eq.(\ref{difference}), one were to set $\delta_\xi \h\gamma_{ab} = \Lie_\xi \h\gamma_{ab}$  --as was done in some of the early work-- then the supermomentum flux would not have the soft term.) Finally, there is some recent discussion on whether there can be radiation of angular momentum without radiation of energy in the context of the post-Minkowskian approximation \cite{Damour:2020tta,Veneziano2022,Manohar:2022dea,Riva:2023xxm}. It follows immediately from Eqs~(\ref{H3}) and (\ref{H4}) that in our framework the answer to this question is in the negative for full nonlinear general relativity. Furthermore, as we will see in section \ref{s3.3}, the angular momentum charge that descends from this flux is the only one among current candidates that has all the physically viable properties.

3. In this section we focused on finite sub-regions $\Rh$ bounded by two 2-sphere cross sections of $\scrip$. To incorporate full $\scrip$, one has to impose suitable boundary conditions on the connections $\{\hD\}$ in the far future and past (i.e., as we move to $i^+$ and $i^\circ$ along $\scrip$). A natural strategy is to consider $C^\infty$ fields and incorporate the appropriate fall-off conditions by endowing the phase space $\psrad$ with the topology of a Fr\'echet space \cite{aams}. Then, the infinitesimal canonical transformations $\delta_\xi$ can be defined on the full phase space since the space of  $C^\infty$ fields on $\scrip$ is preserved by the operation of taking Lie derivatives. Also, the finite canonical transformations generated by the BMS vector fields are also well-defined on entire $\psrad$. We could avoid the use of rather complicated Fr\'echet spaces because fall-off conditions are not needed on finite regions $\Rh$. Reciprocally, since the surface terms that result from integration by parts do not vanish for finite regions (since there are no fall-off conditions at the boundary of $\Rh$), results would not have been stronger had we used Fr\'echet spaces.

\subsection{BMS Charges}
\label{s3.3}

In non-gravitational field theories on given space-times --such as the scalar field theory of section \ref{s2}-- one only has  3-dimensional flux integrals, $F_\xi$, associated with Killing and BMS symmetries $\xi^a$. For the gravitational field, on the other hand, one can also define charges $Q_\xi$ that are 2-sphere integrals. Now, in section \ref{s3.2} we obtained expressions of fluxes using phase spaces $\psRh$ of radiative degrees of freedom of the gravitational field. The radiative modes can be encoded in connections $\{\hD\}$ defined intrinsically on the 3-manifold $\scrip$ equipped with pairs $(\hq_{ab}, \hn^a)$, without reference to 4-dimensional space-times. To discuss associated charges, on the other hand, we need to return to conformal completions $(\hM, \hg_{ab})$ of asymptotically flat space-times $(M,\, g_{ab})$ and use fields that capture the Coulombic information that is \emph{not} registered in $\psRh$. 

We will assume that the completions are smooth (say $C^4$) so that the asymptotic Weyl curvature $\h{K}_{abcd} = \Omega^{-1} \h{C}_{abcd}$ is $C^1$ and we can use the Bianchi identities it satisfies.%
\footnote{This requirement can be weakened so that `peeling' holds only for the Newman-Penrose Weyl components $\Psi_4^\circ,\, \Psi_3^\circ$\, and $\Psi_2^\circ$ and for the ${}_{1}Y_{1,m}$ part of $\Psi_1^\circ$. Bieri \cite{bieri2023radiation} has shown that this weaker peeling holds for a large class of initial data for vacuum solutions that are more general than those considered in \cite{dcsk,pced} in that the mass aspect is now allowed to be `anisotropic'. The BMS charges discussed in this subsection are well-defined under this weaker peeling. } 
Given Einstein's equations, these identities link the radiative degrees of freedom, used so far, with the Coulombic degrees that enter the expression of charges. Thus, to discuss charges, we need to widen our arena by switching from  $\scrip$ as an abstract 3-manifold, to boundaries $\scrip$ of asymptotically flat space-times. The Hamiltonians $H_\xi (\{D\})$ on $\psRh$ will now be regarded as providing fluxes across finite regions $\Rh$ of these boundaries. 

In this framework, charges arise as follows. Recall from section \ref{s3.2} that the fluxes $F_\xi [\Rh]$ across regions $\Rh$ of $\scrip$ arise as Hamiltonians $H_\xi$, and since the expression (\ref{H4}) of $H_\xi$ holds for any open region $\Rh$, for each BMS vector field $\xi^a$ we have a 3-form $\Fxi_{mnp}$ on $\scrip$, representing the local flux of the $\xi^a$-component of the BMS momentum: 
\ba \label{flux1} \Fxi_{mnp}\, &=&\, \frac{1}{16\pi G} \big((\Lie_{\xi} \hD_a - \hD_a \Lie_{\xi})\h\ell_b + 2 \h\ell_{(a} \hD_{b)} \beta \big)\,\h{N}_{cd}\, \hq^{ac}\hq^{bd}\,\, \h\epsilon_{mnp}\nonumber\\ 
&=& \frac{1}{16\pi G} \big(\delta_\xi \hat\gamma_{ab}\big) \,\h{N}_{cd}\, \hq^{ac}\hq^{bd}\,\, \h\epsilon_{mnp} 
\ea
so that the total flux across any region $\Rh$ is given by 
\be \label{flux2} H_\xi[\Rh] =\, F_\xi [\Rh]\, =\, \int_{\Rh} \Fxi_{mnp} \,\,\rmd S^{mnp}\, . \ee
(Here $\h\epsilon_{mnp}$ is the volume 3-form on $\scrip$ in the conformal frame $(\hq_{ab}, \hn^a)$ \cite{gerochrev}.)
Note that although $\scrip$ is now a boundary in $(\hM,\hg_{ab})$, all fields on the right side of (\ref{flux1}) are 3-dimensional, defined intrinsically on it, using the kinematical pairs $(\hq_{ab}, \hn^a)$, and $\hD$ that carries radiative information. The question is whether this flux 3-form is exact. That is, given \emph{any} region $\Rh$ bounded by two cross-sections $S_1$ and $S_2$ of $\scrip$, can one write the flux across $\Rh$ as a difference 
\be \label{balance}  
F_\xi [\Rh]\, = \, \Big(\oint_{S_2} - \oint_{S_1}\Big)\, \Q^{(\xi)}_{ab}\,\, \rmd S^{ab}\ee
between charges associated with $S_1$ and $S_2$, using a charge aspect $\Q^{(\xi)}_{ab}$? This question has been answered in the affirmative \cite{aams,Dray1985}. However, as explained above, these charge 2-forms involve `Coulombic fields' while the flux 3-forms involve only radiative degrees of freedom.

To see this interplay between the radiative and Coulombic fields, let us begin with supertranslations $\xi^a = \super\,\hn^a$. Consider solutions $g_{ab} \in \ps$ and  corresponding (divergence-free) conformal metrics $\hg_{ab}$. The metric $\hg_{ab}$ carries both the radiative and Coulombic information and the two are intertwined via Einstein's equations. Using them (and differential geometric identities) one can systematically `integrate' the flux 3-form\, $\hat\F^{(\super)}_{mnp}$\, step by step and display it as the exterior derivative of a locally defined 2-form $\mathcal{Q}^{(\super)}_{{}_{np}}$ on $\scrip$ \cite{aams}:
\be \F^{(\super)}{}_{mnp}\, =\, (\rmd\, \Q^{(\super)})_{mnp},\quad  {\rm where}\ee
\be \label{Q1}{\Q}^{(\super)}_{{}_{np}} = -\,\frac{1}{8\pi G} \Big[\, \super\,\h{K}^{ab} \h\ell_b \,+\, (\super\hD_b \h\ell_c + \h\ell_b \hD_c \super)\, \h{N}_{de}\, \hq^{ce}\,\hq^{d [a}\, \hn^{b]}\Big]\, \h\epsilon_{anp}\,  \ee
and $\h{K}^{ab}\, := \Omega^{-1}\, \h{K}^{acbd} \hn_c  \hn_d$ is the `electric part' of the asymptotic Weyl curvature. The component $\h{K}^{ab}\h\ell_a \h\ell_b$ that contributes to the integral is denoted by ${\rm Re} \Psi_2^\circ$ in the Newman-Penrose framework. It is `Coulombic'; the connections $\{\hD \}\in \psRh$ have no knowledge of it. This is why the `integration procedure'\, requires access to full $\hg_{ab}$ and its derivatives at $\scrip$, and not just $\hD$ and its curvature.
 
Since $\hat\F^{(\super)} = \rmd {\Q}^{(\super)}$, the desired balance law (\ref{balance}) is satisfied for  supertranslations. The 2-sphere charge integral 
\be Q_{\super} [S] = \oint_S {\Q}^{(\super)}_{{}_{np}}\, \rmd S^{np} \ee
is the component of the supermomentum charge corresponding to $\xi^a = \super \hn^a$. It is conformally invariant, and does not depend on the choice of the inverse metric $\hq^{ab}$. Two features of the charge aspect are noteworthy. 
First, although $\h\ell_a$ appears in the expression of $\mathcal{Q}^{(\super)}_{{}_{np}}$, as in (\ref{flux1}) it can be \emph{any} 1-form satisfying $\hn^a\h\ell_a = -1$; it is \emph{not} tied to the cross-section on which the charge integral is evaluated. The charge $Q_{\super} [S]$ is independent of the choice of $\h\ell_a$. Thus, for any choice of $\h\ell_a$ satisfying $\hn^a\h\ell_a =-1$, $\mathcal{Q}^{(\super)}_{{}_{np}}$ is a \emph{local} 2-form on $\scrip$. Second, while  ${\Q}^{(\super)}_{{}_{np}}$ contains `Coulombic information' through $\h{K}^{ab} \h\ell_a\h\ell_b$, this information disappears once we take the exterior derivative of ${\Q}^{(\super)}_{{}_{np}}$; the resulting flux 3-form contains purely radiative information! \vskip0.1cm

Next, let us address the question of uniqueness of the supermomentum charge $Q_{\super} [S]$. By inspection, the charge aspect $\Q^{(\super)}_{np}$ of (\ref{Q1}) satisfies the following viability criteria:\vskip0.1cm
\noindent (i) it is a 2-form whose exterior derivative is the flux 3-form;\\
(ii) it is locally constructed from fields at $\scrip$, is linear in $\super$, and has zero conformal weight; \\
(iii) in Minkowski space, it leads to zero supermomenta on any cross-section $S$ of $\scrip$. \vskip0.1cm 
\noindent Is there another candidate $\tilde{\Q}^{(\super)}_{{}_{ab}}$ that also satisfies them? Suppose there is. 
Then the difference \smash{$(\Delta{\Q})_{ab}{}^{(\super)}:= \tilde{\Q}^{(\super)}_{ab} - {\Q}^{(\super)}_{ab}$} also satisfies (ii) and (iii), and (i) implies that it is a closed 2-form on $\scrip$ of \emph{every} asymptotically flat solution. In particular, then, in any given space-time, the value of the difference between two charges,\, $\Delta Q_{\super}:=\oint_S \Delta\Q^{(\super)}_{ab} \rmd S^{ab}$,\, does not depend on the cross-section $S$. 
Let us introduce a foliation of $\scrip$ by $u={\rm const}$ cuts. Since  any 2-form can be expanded in a coordinate basis, we can expand\,  $(\Delta\Q)^{(\super)}_{ab}$\, as\, $(\Delta\Q)^{(\super)}_{ab} = \super (f_1 \h\epsilon_{ab} + f_2\, v_{[a} \hD_{b]} u )$\, for some 1-form $v_a$. Let us evaluate the charge on a $u\! =\! u_\circ$ 2-sphere $S_\circ$. Since the pull-back of $\hD_b u$ to $S_{\circ}$ vanishes, we have $\Delta Q_{\super} = \oint_{S_\circ} \super\, f_1 \epsilon_{ab}$ where,\, by condition (ii),\, $f_1$ is a locally defined field at $S_\circ$. Since $(\Delta Q)^{(\super)}_\xi$ is time independent for all supertranslations $\super$, it follows that $\Lie_{\hn} f_1 =0$ for all space-times in $\ps$. Therefore $f_1$ cannot depend on local dynamical variables; it must be constructed from the universally available kinematical fields. It follows that the value of $\Delta Q_{\super}$  is the same for all space-times under consideration, including Minkowski space-time. Finally, condition (iii) implies that this value is zero in Minkowski space-time, hence  $\Delta Q_{\super} =0$. We conclude that the supermomentum charges defined by $\tilde{\Q}^{(\super)}_{{}_{np}}$ are the same as those defined by our $\Q^{(\super)}_{np}$.

Let us now consider a general BMS vector field $\xi^a$. The charge aspect $\Q^{(\xi)}_{np}$ can also be obtained by `integrating' the flux $F_\xi [\Rh]$ \cite{Dray1985} but, as we discuss below, there is a conceptual difference. Let us fix a cross-section $S$ of $\scrip$. Then any BMS vector field $\xi^a$ can be decomposed as $\xi^a = \super \hn^a + \zeta^a$, where $\super\hn^a$ is a supertranslation and $\zeta^a$ is tangential to $S$, and is thus a generator of a Lorentz transformation. (Note that this split refers to a single cross-section, not to a foliation of $\scrip$.) Since we already have the expression of the charge aspect for supertranslations, let us focus on $\zeta^a$. It turns out that, in contrast to $\mathcal{Q}^{(\super)}_{np}$, the definition of $\mathcal{Q}^{(\zeta)}_{np}$ requires us to tie  the choice of the 1-form $\h\ell_a$ to the surface $S$ on which the charge is to be computed. Given a cross-section $S$, let us choose $\h\ell_a$ to be the normal to that cross-section satisfying $\hn^a \h\ell_a = -1$ as before. Then, the shear of $S$ is given by $\h{\sigma}_{ab} = {\rm TF}\, \h{q}_a{}^c \h{q}_b{}^d\, \hD_{(c} \h\ell_{d)}$, where $\h{q}_a{}^c$ is the projection operator on the 2-sphere $S$. Finally, let us introduce the 1-form $ \h{\mathcal{K}}_d = \h{K}^{abc}{}_d \, \h\ell_a\,\hn_b \,\h\ell_c$ using the asymptotic Weyl tensor. These fields are used to define the charge aspect satisfying the balance law (\ref{balance}) \cite{Dray1984,Dray1985}:  
\ba \label{Q2} Q_{\zeta} [S] &=& - \frac{1}{8\pi G}\, \oint_S \zeta^a\Big[\h{\cal K}_{a} + {\h\sigma   }_{ab}\, 
{{}^2\!\hD}_c{\h\sigma}^{bc}\,+\,\textstyle{\f{1}{4}} \, {{}^2\!\hD}_a ({\h\sigma}_{bc} {\h\sigma}^{bc}) \Big] \, \h\epsilon_{np} \, \rmd {S}^{np} \nonumber\\
&\equiv& - \frac{1}{8\pi G}\, \oint_S \Q^{(\zeta)}_{{}_{np}}\, \rmd {S}^{np} \ea 
where indices are raised and lowered using the intrinsic metric on $S$ and\, ${{}^2\!\hD}$ 
is the 2-sphere derivative operator compatible with this metric. Combining Eqs.~(\ref{Q1}) and (\ref{Q2}), we now have the expression of the charge $Q_{\xi}[S]$ associated with a general BMS vector field $\xi^a$ at any 2-sphere cross-section $S$ of $\scrip$,
\be \label{Q3} Q_\xi[S] = Q_{\super} [S] + Q_{\zeta} [S]\, ,  \ee
which satisfies the balance law (\ref{balance}) with the flux given by (\ref{flux1}) and (\ref{flux2}). Note, however, that, in contrast to the supermomentum charge aspect $\Q^{(\super)}_{{}_{np}}$,\, the angular momentum charge aspect $\Q^{(\zeta)}_{{}_{np}}$  cannot be specified once and for all, but only once the cross section has been specified.
Therefore verification of the balance law (\ref{balance}) is a bit more involved (see Remark 1 below).  
\goodbreak

\emph{Remarks}: 

1. The expression of the supermomentum charge aspect (\ref{Q1}) was first postulated by Geroch in \cite{gerochrev}, motivated by its properties. Subsequently, fluxes (\ref{H4}) for all BMS vector fields were derived using the phase space of radiative modes and the supermomentum charge was obtained by a step by step integration of the flux expression in \cite{aams}. The angular momentum charge was first was postulated by Dray and Streubel in \cite{Dray1984}) using considerations from Twistor theory. Dray \cite{Dray1984} then showed that it arises from the flux expression (\ref{H4}) (that was already available in the literature \cite{aams}) using the following procedure. The boundary cross-sections $S_1$ and $S_2$ bounding any $\Rh$ are related by a supertranslation. The generator of this supertranslation was used to foliate the given $\Rh$. The charge aspect $\Q^{(\xi)}_{np}$ was introduced on $\Rh$ using the normal $\h\ell_a$ to the leaves of this foliation, and the balance law (\ref{balance}) was verified using the Newman-Penrose formalism. More recent treatments \cite{Barnich:2011mi,Flanagan:2015pxa,Kesavan,Grant:2021sxk,Odak:2022ndm} also introduce a foliation and the associated null normal $\ell_a$, but do not require the extra structure of the Newman-Penrose formalism. 

We presented a self-contained summary of these results that is conceptually more complete in that it brings out the distinction between fluxes and charges, as well as the subtle difference between the nature of charge 2-forms for supermomenta versus angular momenta. Fluxes require only radiative degrees of freedom that are encoded \emph{intrinsically} in $\Rh$. For them, the 4-metric and the full field equations are excess baggage. Charges on the other hand refer to Coulombic aspects that are intertwined with the radiative aspects through field equations; one needs to step out of $\psRh$ and use information from full $\ps$. Similarly, while the supermomentum charge aspect can be specified once and for all on all of $\scrip$, the angular momentum charge aspect is tied to the cross-sections under consideration. This difference has not been emphasized in the literature. Finally, we also took this opportunity to correct minor errors in some of the older literature.

2. All contemporary literature uses the expression (\ref{Q1}) of supermomentum. For angular momentum, on the other hand, alternate candidates have also been proposed. Proposals in which the charges constitute BMS momentum --i.e. is linear in the BMS generators $\xi^a$, as in (\ref{Q3})-- are summarized in \cite{Compere:2019gft,Elhashash:2021iev}. 
The alternate expressions have two types of drawbacks. First, in general axisymmetric space-times with non-vanishing $\h{N}_{ab}$, they lead to non-zero angular momentum fluxes around the symmetry axis. Second, they do not admit a local flux, whence their angular momentum charge does not change continuously with continuous deformations of the cross-section $S$ \cite{Chen:2022fbu}. There is also a more recent candidate angular momentum charge that stemmed from a well-developed `quasi-local charges' framework \cite{Chen:2021szm}. However, this strategy does not provide a linear map from all BMS generators to reals.
\footnote{Any definition of angular momentum that is free of supertranslation ambiguity cannot be a linear map from all BMS generators to the reals. It is sometimes argued that such a definition of angular momentum is needed in order to obtain the standard transformation law of special relativity for
a boosted stationary solution. 
This  expectation is incorrect: Our BMS charge (\ref{Q3}) does have the correct transformation property. Indeed, if the boost is implemented in a (coordinate) invariant manner \eqref{Q3} has this property in any stationary space-time \cite{Ashtekar:1979iaf}.}
Rather, on any given cross-section $S$ of $\scrip$, the procedure selects a rotation subgroup of the BMS group \emph{using physical fields at that cross-section} and associates angular momentum charges with them. But in axisymmetric space-times, this procedure generically excludes the rotations generated by the \emph{exact Killing field} from its rotation subgroup, whence these angular momentum charges do not include the generally accepted notion physical angular momentum. Among expressions of fluxes and charges available in the literature, \eqref{Q3} is the only one that is free from  all these limitations.

3. Finally, note that in the passage from fluxes to charges, we used the fact that $\ps$ is `sufficiently rich'. For example, if we had restricted ourselves to its `non-radiative' sector $\ps^\prime$ consisting \emph{only} of solutions $g_{ab}\in\ps$ for which ${}^\star\!\h{K}^{ab}$ vanishes on $\scrip$, all flux 3-forms $\Fxi_{mnp}$ would have vanished although, as is clear from their expressions (\ref{Q1}) and (\ref{Q2}), charges do not vanish on $\ps^\prime$. If we had restricted ourselves to $\ps^\prime$ from the start, our procedure would have yielded the correct --namely zero-- fluxes. However, the uniqueness argument for charges, given above, would not have sufficed because, while the only fields at $\scrip$ that are time independent on the \emph{full} phase space $\ps$ are kinematical, there are physical fields (such as ${\rm Re} \Psi_2^\circ$) that are time-independent on entire $\ps^\prime$. They provide non-trivial candidate charges on $\ps^\prime$ that are compatible with zero fluxes.  More physical inputs would have been necessary to arrive the correct charge expressions. While this is a rather trivial observation, it is of direct relevance to our discussion of charges on $\Delta$ in the next section since all fields on  $\Delta$ are non-radiative.

\section{Black Hole (and Cosmological) Horizons $\Delta$}
\label{s4}

Structure of black hole and cosmological horizons $\Delta$ was discussed in detail in the companion paper \cite{aass1}. We will now apply the Hamiltonian framework introduced in section \ref{s2} to these horizons $\Delta$, drawing heavily on \cite{aass1}.

Since these horizons $\Delta$ are null 3-manifolds in physical space-times, no conformal completion is involved. In this case, $\ps$ consists of solutions $g_{ab}$ to Einstein's equations that admit a  WIH \,$\Delta$ as internal boundary. The detailed construction of this $\ps$, spelled out in \cite{akkl1}, can be abbreviated as follows. Let $M$ be a 4-manifold with an internal boundary $\Delta$ with topology $\mathbb{S}^2\times\mathbb{R}$. Equip $\Delta$ with the universal structure of a  WIH: a 3-parameter family of pairs  of fields $(\qo_{ab},\, [\ello^a])$, where  $\qo_{ab}$ is a unit, round, 2-sphere metric, and $[\ello^a]$ is an equivalence class of vector fields (along the $\mathbb{R}$ direction, where\, $\ello^a \approx c \ello^a$\, for any positive constant $c$), such that any two pairs are related by $(\qo^\prime_{ab},\, [\ello^{\prime\,a}]) =(\alphao^2\qo_{ab},\, [\alphao^{-1}\ello^a])$. ($\alphao$ is constrained to ensure that $\qo_{ab}$ and $\qo^\prime_{ab}$ are both unit round metrics; see Eq. (3.2) of \cite{aass1}.) These fields will provide the kinematical structure on $\Delta$. The phase space $\ps$ now consist of solutions $g_{ab}$ on $M$ in which:
\footnote{The second condition in (i) can always be imposed by a gauge transformation on $g_{ab}$,\, i.e., using a diffeomorphism on $M$ that is identity on $\Delta$ \cite{cfp}. It simplifies the subsequent analysis \cite{akkl1}.}\vskip0.1cm

\begin{enumerate}[label=\upshape(\roman*)]
\item\label{lformuniversal} $\Delta$ is a WIH and, given an $\ell^a$ from the canonical equivalence class $[\ell^a]$ induced by $g_{ab}$ on $\Delta$, the 1-form $\ell_a := g_{ab} \ell^a$ is the same for all $g_{ab} \in \ps$; and,
 
\item The metric $q_{ab}$ and the canonical null normal $[\ell^a]$ induced on $\Delta$ by any $g_{ab} \in \ps$  are conformally related to any given pair $(\qo_{ab}, \, [\ello^a])$ in the kinematical structure: $q_{ab} = \psio^{-2} \qo_{ab}$ and $[\ell^a] = [\psio \ello^a]$, for some positive function $\psio$ satisfying $\Lie_{\ello}\, \psio =0$.
\end{enumerate}

Our next task is to extract from $\ps$ the phase space $\psDelta$ tailored to degrees of freedom that reside on $\Delta$. 
Recall that the metric $g_{ab}$ induces on $\Delta$ a triplet of fields, $q_{ab}, [\ell^a]$, and $D$, that constitute the WIH geometry (see section II.A of \cite{aass1}). As at $\scrip$, they capture the information in $g_{ab}$ that resides on $\Delta$. Recall that at $\scrip$ the fields $(\h{q}_{ab}, \h{n}^a)$ are part of the kinematical structure, and $\hD$ carries the physical information that varies from one space-time to another. By contrast, since $\Delta$ is a sub-manifold of the physical space-time, rather than of a conformal completion thereof, now there is physical information in $q_{ab}$ as well; the kinematical structure is confined to the fields $(\qo_{ab}, [\ello^a])$. As at $\scrip$, it is convenient to isolate the freely specifiable data on $\Delta$ by fixing fiducial fields. Let us fix a kinematical pair $(\qo_{ab},\, \ello^a)$. Then the fields $(q_{ab},\, \ell^a)$ induced by the physical metric $g_{ab}$ are completely determined by the positive function $\psio$. As discussed in section II of \cite{aass1}, $D$ is completely determined by the 1-form $\omegao_a$ and a symmetric tensor field $\co_{ab}$ that is transverse to $\ello^a$.
\footnote{$\omegao_a$ is defined by  $D_a \ello^b = \omegao_a \ello^b$, and $\co_{ab}$ is obtained by replacing $n_a$ in Eq. (2.9) of \cite{aass1} by $\no_a = \psio\, n_a$.}
Thus, a point of $\psDelta$ can be conveniently labeled by the triplet $(\psio, \omegao_a, \co_{ab})$ of fields on $\Delta$ that are all Lie-dragged by $\ello^a$ and, in addition, $\omegao_a$ and $\co_{ab}$ are transverse to $\ello^a$. Thus, $\psDelta$ admits a single global chart. While $\ps$ is a highly non-linear space, the phase space $\psDelta$ of degrees of freedom that reside on $\Delta$ is essentially linear, just as $\psrad$ is at $\scrip$. \vskip0.1cm

However, as noted in section \ref{s2} of \cite{aass1}, in striking contrast to fields $\gamma_{ab}$ labelling points $\{\hD\}$ of $\psrad$, the triplet of fields $(\psio, \omegao_a, \co_{ab})$ labeling points of $\psDelta$ are non-dynamical: they are `time independent' and therefore carry 2-dimensional --rather than 3-dimensional-- degrees of freedom. Mathematically they constitute the `corner data' and physically they carry `Coulombic' rather than `radiative' information. This fact has a key consequence on the phase space structure. Recall that the symplectic current $\J_{mnp}$ of $\ps$ is evaluated at any $g_{ab}$ and depends linearly on two tangent vectors $h_{ab}, \, h^\prime_{ab}$ at this $g_{ab}$. As usual, the symplectic current on $\psDelta$ is obtained by pulling back $\J_{mnp}$ of $\ps$ to the 3-manifold $\Delta$. Let us restrict ourselves to $g_{ab} \in \psDelta$ and to tangent vectors that preserve the properties that these $g_{ab}$ have to satisfy. Then, one finds that the pull-back of $\J_{mnp}\!\!\mid\!\!_{{}_{\psDelta}}$ to $\Delta$ vanishes identically \cite{akkl2,cfp}! This is in striking contrast to the situation at $\scrip$ but just what one would physically expect, given that $\psDelta$ does not have  any fields that carry 3-dimensional (radiative) degrees of freedom.

Finally, as at $\scrip$, we can introduce candidate phase spaces $\psRDelta$ associated with regions $\R$ of $\Delta$. They are obtained by restricting the triplet $(\psio,\, \omegao_a, \co_{ab})$ to $\R$. Obviously, the restriction $\J_{mnp}\!\!\mid\!\!_{{}_{\psRDelta}}$ of the symplectic current vanishes identically. It then immediately follows that fluxes $F_\xi$ associated with all symmetry vector fields $\xi^a$ also vanish. Therefore, it is no longer necessary to take the additional step of equipping $\psDelta$ with a topology. To summarize, the strategy of using the Hamiltonian framework associated with finite regions $\R$ of $\Delta$ again yields the flux 3-forms $\F^{(\xi)}_{mnp}$,\, but they all vanish. This is exactly what one would expect from physical properties of $\Delta$.\smallskip

Our next task is to find expressions of charges. At $\scrip$ we could just begin with the flux 3-forms $\Fxi_{mnp}$ and use field equations and Bianchi identities to express them as exterior derivatives of the charge aspects $\Q^{(\xi)}_{np}$. As remarked at the end of section \ref{s3.3}, this is possible at $\scrip$ because the phase space was sufficiently rich; had we restricted ourselves to the \emph{non-radiative} subspace $\ps^\prime$ of $\ps$ from the beginning, $\Fxi_{mnp}$ would have been identically zero and a priori there would be many distinct candidates for the charge aspect $\Q^{(\xi)}_{np}$. As in any other approach, one would have needed additional inputs to find the physically appropriate charges. 

At $\Delta$ we face the same situation. Our task is to find $\Q^{(\xi)}_{np}$ that depend locally on fields induced on $\Delta$ by $g_{ab}$, and are \emph{closed for all} $g_{ab} \in \ps$ (so that all fluxes vanish, as desired). It is clear that the fields that enter the expression of $\Q^{(\xi)}_{np}$ have to be time-independent on full $\ps$ associated with $\Delta$. At $\scrip$, such fields have to be kinematical. By contrast now   \emph{every} $g_{ab} \in \ps$ provides us with time-independent fields, --such as $q_{ab}$ and $\omega_a$-- that carry \emph{physical information}. In fact they carry precisely the Coulombic information that is needed to introduce charges.  And, since the fields are time independent, the charges they define would be automatically conserved, leading to zero fluxes as desired. One would have to impose additional physical criteria in order to single out charge aspects $\Q^{(\xi)}_{np}$. 

Instead, we will use another strategy that also serves to make contact with the Wald-Zoupas one: we will use the phase spaces $\psSigma$ associated with \emph{partial} Cauchy surfaces $\Sigma$ that join cross-sections $S$ of $\Delta$ to $i^\circ$. This strategy is suggested by the fact that while the cross-sections $S$ of $\Delta$ are boundaries of $\Sigma$ --just as they are of regions $\R$ within $\Delta$-- the flux of the symplectic current does not vanish across $\Sigma$. Let $X^a$ be a vector field on $M$ that is tangential to $\Delta$ and generates diffeomorphisms that map each $g_{ab} \in \ps$ to a $g^\prime_{ab} \in \ps$. In order to avoid specifying boundary conditions at $i^{\circ}$, let us also assume that $X^a$ vanishes outside a spatially compact world-tube in $M$. Then, as discussed in section III of \cite{aass1}, in any conformal frame $(\qo_{ab}, \ello^a)$ from the universal structure on $\Delta$, the restriction $\xi^a$ of $X^a$ to $\Delta$ has the form 
\be \label{xi3} \xi^a\, \=\, V^a_{\xi} + H^a_{\xi} \ee
where the vertical and horizontal components of $\xi^a$ are given by
\be V^a = \big((\varpio + \betao)\vo + \mathring{\mathfrak{s}})\big) \ello^a, \qquad {\rm and} \qquad H^a_\xi = \epsilono^{ab} \Do_b \chio\, - \, \qo^{ab} \Do_b \betao\, . \ee
Here: (i) $\vo$ is an affine parameter of $\ello^a$ and  $\epsilono^{ab},\, \qo^{ab}$ are the inverses of the area 2-form and the metric on the $\vo = {\rm const}$ cross-sections, respectively;\,\, (ii) $\mathring{\mathfrak{s}}(\vartheta,\varphi)$ is a general function on the 2-sphere of null generators of $\Delta$ and $\mathring{\mathfrak{s}} \ello^a$ represents a supertranslation; \, (iii) $\chio(\theta,\varphi)$ and $\betao(\vartheta,\varphi)$ are both linear combinations of first three spherical harmonics defined by $\qo_{ab}$ and $\epsilono^{ab} \Do_b \chio\,$ and $(\betao\vo \ello^a - \qo^{ab} \Do_b \betao\,)$ are generators of rotations and boosts; and, (iv) $\varpio$ is a constant and  $\varpio\,\ello^a$ is a generator of dilations (a symmetry of $\Delta$ that has no counterpart at $\scrip$).

Using the fact that $g_{ab}$ satisfies Einstein's equations on $M$, one can show that for any $g_{ab} \in \ps$ and tangent vectors $\delta$ and $\delta_X$\, \cite{akkl2},  
\be \label{variation} \bfomegaSigma\!\mid_{g}\, (\delta_{X},\, \delta) =  \frac{1}{8\pi G}\,\,\delta \oint_S \big(\varpio + H^a_\xi\, \omega_a\big)\,\epsilon_{np}\, \rmd S^{np}\, =:\,  \frac{1}{8\pi G}\,\,\delta \oint_S \Q^{(\xi)}_{np}\, \rmd S^{np} \,. \ee
Note that: (i) the right side depends \emph{only on} the restriction $\xi^a$ of $X^a$ to $\Delta$, and, (ii) all other fields in the integrand are also evaluated on $\Delta$, and are furthermore time-independent. Therefore, it follows that the 2-form integrand $\Q^{(\xi)}_{np}\, =\, \big(\varpio + H^a_\xi\, \omega_a\big)\,\epsilon_{np}$  is closed, whence we (trivially) have\, $d\Q^{\xi} = \F^{\xi}$. Note also that $\Q^{(\xi)}_{np}$ is built from fields that are intrinsically defined and locally constructed from $g_{ab}$, and linear in the symmetry generator. Therefore we are led to define charges on $\Delta$ as:
\be \label{QDelta} Q_{\xi}[S] = \frac{1}{8\pi G} \oint_S \big(\varpio + H^a_\xi\,\omega_a\big)\,\epsilon_{np}\, \rmd S^{np} \ee 
for all $\Delta$-symmetry  generators $\xi^a$. A priori, the Eq.~(\ref{variation}) determines the $Q_\xi [S]$ only up to the addition of a constant. We have eliminated this freedom by demanding that, along the 1-parameter family of Schwarzschild WIHs, all charges should vanish in the limit in which area of the WIH vanishes, following \cite{cfp}.

Note that the supertranslation descriptor $\mathring{\mathfrak{s}}$ does not appear on the right side, whence all supermomentum charges vanish on $\Delta$. This may seem surprising at first because one might expect that the static Killing field in the Schwarzschild space-time, for example, would be a supertranslation on $\Delta$. However, that is not correct; as we pointed out in section \ref{s3.1} of \cite{aass1}, it is a dilation field on $\Delta$ and the dilation descriptor $\varpio$ does appear, whence the dilation charge is non-zero. Similarly, in non-extremal Kerr space-times, the linear combination of the two Killing fields that is normal to the horizon is a dilation. In the extremal case, it \emph{is} a supertranslation, but the linear combination of the asymptotic time translation and the rotational Killing field that is null at the horizon is such that the corresponding charge $Q_\xi$ is the linear combination of the mass and angular momentum that vanishes, in agreement with our finding that all supermomentum charges vanish on $\Delta$. More generally, this definition of charges passes a number of non-trivial physical criteria discussed in \cite{akkl2}. It also agrees with the charges obtained by the extension of the Wald-Zoupas procedure given in \cite{cfp} when restricted to a WIH. However, our procedure does not need a preferred symplectic potential: the detailed examination shows that it bypasses this step by taking advantage of the fact that fluxes across $\Delta$ vanish. 
\\
\goodbreak
\emph{Remarks:}

1. Eq.~(\ref{variation}) implies that $Q_\xi [S]$ is the Hamiltonian generating the canonical transformation $\delta_X$, induced by the vector field $X^a$ on $\psSigma$. As discussed in Appendix \ref{a1}, charges defined at $\scrip$ by contrast, do not admit the analogous interpretation on full phase spaces $\psRh$. This difference arises because while the fluxes $\Fxi_{mnp}$ are generically non-zero across $\scrip$, the fluxes $\F^{(\xi)}_{mnp}$ vanish identically on $\Delta$.

2. Note that while the left side of Eq.~(\ref{variation}) features the vector field $X^a$ in $M$, the right side is sensitive only to its restriction $\xi^a$ to $\Delta$. In the calculation, to begin with the right side does feature a directional derivative of $X^a$ that is transversal to $\Delta$. However, because $X^a$ has to preserve property \ref{lformuniversal} in the definition of $\ps$, this term is constrained: it is determined by the the restriction $\xi^a$ of $X^a$ to $\Delta$. That is, $X^a$ can be \emph{any} space-time vector field that preserves the universal structure on $\Delta$ that is common to all $g_{ab} \in \ps$. This is why, the charge $Q_{\xi}[S]$ is insensitive to the extension $X^a$ of $\xi^a$ away from $\Delta$; it depends only on $\xi^a$ just as one would physically expect. For details on this subtlety, see Appendix B of \cite{akkl2}, and \cite{cfp,Odak:2023pga}. 
3. The WZ strategy was applied in \cite{cfp} to an arbitrary null hypersurface, identifying a unique symplectic potential which is covariant and vanishes on NEHs in vacuum. One can apply the new strategy described in sections \ref{s2}- \ref{s4} to this case as well. Remarkably, the analogue of the norm \eqref{norm2} in this case identifies again the unique WZ symplectic potential, as one can easily check. Therefore the new strategy and the WZ one give consistent results in different physical settings such as $\scrip$ and arbitrary null hypersurfaces.

\section{Discussion}
\label{s5}

The fact that null infinity is a WIH seems very surprising at first because WIHs are commonly associated with black hole (and cosmological) horizons $\Delta$ in equilibrium. Indeed, $\scrip$ and $\Delta$ have almost the opposite connotations. $\scrip$ lies in the asymptotic, weak field region, while $\Delta$ lies in a strong curvature region. $\scrip$ is the arena for discussing gravitational waves, while $\Delta$ is generally used to discuss the Coulombic properties of black holes in equilibrium. The companion paper \cite{aass1} showed that, in spite of these striking contrasts, they share a number of geometrical properties. In particular, one can systematically arrive at the BMS group $\B$ at $\scrip$ starting from the symmetry group $\G$ of WIHs. In this paper we continued our exploration of unity that underlies apparent diversity.  We showed that, while fluxes associated with the BMS group $\B$ of $\scrip$ have a rich structure and those associated with  $\G$ at $\Delta$ simply vanish, these diverse conclusions arise from the same conceptual setting.

To this end, we introduced a new Hamiltonian framework that could be useful also in other contexts. (For example, as pointed out in the Introduction, it is well-suited to future space-like infinity $\scrip$ in asymptotically de Sitter space-times.) It has three novel features:\\
(1) It introduces Hamiltonian methods for degrees of freedom that reside in \emph{finite sub-regions} of suitably chosen 3-dimensional surfaces. \\
(2) Since $\scrip$ and $\Delta$ are null surfaces, it is possible to extract the \emph{unconstrained} degrees of freedom of general relativity that lie in their sub-regions $\Rh$ and $\R$. Consequently, while the covariant phase space $\ps$ is highly non-linear, the local phase spaces $\psRh$ and $\psR$ are essentially \emph{linear}.
(3) Fluxes $F_\xi [\Rh]$ associated with the BMS symmetries $\xi^a$ emerge as Hamiltonians generating canonical transformations on $\psRh$, induced by the action of $\xi^a$ on $\scrip$.
\smallskip

Most discussions of charges and fluxes associated with the BMS symmetries of general relativity is based on the covariant phase space $\ps$. However one typically works only formally with this infinite dimensional nonlinear space; its underlying manifold structure is rarely made precise. In the case of  $\psRh$ on the other hand it is relatively straightforward to spell out their manifold structure and topology. It is instructive to compare and contrast $\ps$ with the phase space of $\psRh$ of local degrees of freedom.
While the covariant phase space $\ps$ of general relativity carries only a (formally defined) pre-symplectic structure, $\psRh$ is endowed with a continuous, weakly non-degenerate symplectic structure $\bfomegaRh$. Non-degeneracy is a consequence of the fact that, in contrast to the covariant phase space $\ps$, there are no gauge degrees of freedom in $\psRh$. Topology on $\psRh$ and continuity of $\bfomegaRh$ provide a degree of mathematical control in the following sense. 
The action of any BMS vector field $\xi^a$ naturally leads to a Hamiltonian vector field $\delta_\xi$ on a dense subspace $\DRhp$ of $\psRh$: it satisfies \smash{$\bfomegaRh(\delta_\xi, \delta) = \delta H_\xi$} on $\DRhp$.%
\footnote{If the symmetry vector field $\xi^a$ is tangential to the boundary of $\Rh$, the action can be integrated to finite symplectic transformations; if not, we only have a densely defined Hamiltonian vector field. There is an analogue situation in quantum theory: generators of space-time symmetries are represented by \emph{densely defined} operators on the Hilbert space of quantum states. If the operator is self-adjoint, the action can be integrated to unitary transformations. Sometimes --as in the case of the infinitesimal action of translations on the Hilbert space of a particle on the half line-- the operator is only symmetric and not self-adjoint, and we only have infinitesimal motions.}
(As usual, there is freedom to add a constant to $H_\xi$ which is eliminated by requiring that $H_\xi (\{ \hD \})$ should vanish at points $\{\hD \}$ of $\psRh$ for which \smash{$N_{ab}=0$}.) Since the symplectic structure on $\ps$ has infinitely many degenerate directions, there is a corresponding gauge ambiguity in the infinitesimal canonical transformation generated by any given Hamiltonian on $\ps$. By contrast, thanks to the non-degeneracy of $\bfomegaRh$ 
this ambiguity disappears and $\delta_\xi$ is the unique Hamiltonian vector field generated by $H_\xi$. 
Finally, $H_\xi$  is continuous on $\DRhp$. Therefore, it can be uniquely extended function all of $\psRh$.  

Since continuity is determined by the choice of topology, it is natural to ask whether our choice is natural. As explained in section \ref{s2}, the choice is initially motivated by examining zero rest-mass scalar fields in asymptotically flat space-times. In that case, the choice of topology is such  that a sequence $\h\phi_n$ of radiation fields in $\psRh$ converges to a field $\h\phi$ if and only if $\h\phi_n$ and their first derivatives converge to $\h\phi$ and its first derivative in an $L^2$ sense on $\Rh$. (This topology is insensitive to the additional structures needed to define the `$L^2$-sense'; it depends only those structures that are naturally available at $\scrip$.) 
This choice of topology leads one to a unique Hamiltonian $H_\xi$ on $\psRh$ that  agrees with the \emph{physically correct flux} $F_\xi [\Rh]$ that is \emph{defined by the stress-energy tensor}:\,  $F_\xi [\Rh] = \int_{\Rh} T_{ab}\, \xi^a \rmd S^b$ for all BMS vector fields $\xi^a$.  Thus, the strategy leads to the physically correct fluxes using Hamiltonian methods, without having to assume the existence  of the stress-energy tensor. It is thus extremely well suited to general relativity where there is no stress-energy tensor for the gravitational field. But topological considerations go through and provide a strategy to define fluxes: since $\psRh$ again consists of radiative modes that reside in $\Rh$, we can equip $\psRh$ with the same topology and calculate Hamiltonians  $H_\xi$ and interpret them as fluxes $F_\xi [\Rh]$ of BMS momenta across $\Rh$. The fact that these fluxes agree with those in the literature reenforces the motivation for choosing this topology. Note that this procedure to arrive at fluxes $F_\xi [\Rh]$ uses (fields and) symmetry vector fields $\xi^a$ that are intrinsically defined on $\scrip$; one does not need to extend them into the space-time interior.

The BMS fluxes $F_\xi [\Rh]$ can be obtained as Hamiltonians generating canonical transformations induced by BMS vector fields on $\psRh$ because the expressions of  $F_\xi [\Rh]$ refer only to radiative modes that are captured in $\psRh$. Charges, on the other hand, are 2-sphere integrals that refer also to the Coulombic information in $\ps$ that is filtered out in the passage to $\psRh$. Therefore, to recover charges from fluxes, one needs to return to $\ps$ and use field equations and Bianchi identities at $\scrip$. One then obtains  the charge 2-forms $\Q^{(\zeta)}_{{}_{np}}$ by `integrating' the flux 3-forms $\Fxi_{mnp}$. As discussed in section \ref{s3.3}, this was in fact the procedure used to arrive at the Dray-Streubel charges \cite{aams,Dray1985}. Thus, in the present approach one first obtains fluxes  $F_\xi [\Rh]$ using radiative phase spaces $\psRh$ and Hamiltonian considerations, and then arrives at charges $Q_\xi [S]$ in a second step. However, this step is also carried out at $\scrip$, without having to extend fields or symmetry generators to the space-time interior. This concludes our summary of the underlying strategy and main results. \smallskip

Next, let us briefly compare and contrast this framework with other frameworks that also discuss null infinity (for details, see Appendix \ref{a1}). \\
\indent (i) Expressions of fluxes and charges satisfy a variety of physical requirements both on $\scrip$ and $\Delta$.  In particular, all fluxes across $\Delta$ vanish, just as one would physically expect. At $\scrip$, the fluxes vanish if the News $\h{N}_{ab}$ vanishes, and,  even when $\h{N}_{ab}$ is non-zero, if the symmetry vector field arises from a Killing field in space-time. All charges vanish in Minkowski space-time, and, the angular momentum charges agree with the Dray-Streubel charges on general space-times. These requirements are not always met in other approaches (see, e.g., \cite{Geroch:1981ut,Chen:2021szm,Chen:2022fbu,Compere:2019gft,Elhashash:2021iev}). \\
\indent (ii) In the discussion of null infinity, the procedure does not need auxiliary structures such `preferred symplectic potentials' that have to be selected in the Wald-Zoupas (WZ) procedure and its extensions \cite{wz,Grant:2021sxk,Odak:2022ndm}. One works with just the symplectic 2-form. The physical fields that feature in the discussion are all intrinsically defined on $\scrip$, using only the region $\Rh$ for fluxes ${F}_\xi [\Rh]$ and 2-spheres $S$ for charges $Q_\xi[S]$. Additional structures that are often used --such as tetrads or coordinates on $\scrip$ (and sometimes also in its neighborhood)--  are not needed. Similarly, one only uses symmetry vector fields $\xi^a$ at $\scrip$ alone. In contrast to other approaches, one does not need prescriptions to extend them to a neighborhood in the space-time interior.\\
\indent (iii) Reciprocally, our approach requires an ingredient that is not needed in other approaches: local phase spaces $\psRh$ with appropriate topology. Thanks to the characteristic initial value problem in general relativity \cite{rendall}, one can extract the degrees of freedom that reside in open regions of null hypersurface and construct local phase spaces with the required topology. The resulting fluxes agree with those obtained using the preferred symplectic 1-form required in the WZ procedure. Thus, there is considerably synergy. However, there are also differences. As remarked above, our procedure can also be used at future space-like infinity $\scrip$ for asymptotically de Sitter space-times, where the preferred symplectic potential satisfying all requirements does not exist \cite{kl2021}. Reciprocally, while the procedure discussed in this paper is well-developed only for general relativity, the WZ procedure offers an avenue to incorporate higher derivative  gravity theories as well. The open issue for our procedure is whether one can introduce the required local phase spaces. This could be difficult because, while in general relativity we relied on results from the characteristic initial value problem, typically this problem has not been studied in higher derivative theories. In the WZ framework the open issue is whether there is a unique symplectic potential with desired properties.
\smallskip

Finally, this unified treatment of black hole (and cosmological) horizons $\Delta$ and null infinity $\scrip$ opens new directions for further research both in classical general relativity and quantum gravity. We will conclude with a few illustrations. For binary black hole mergers, our framework provides an avenue to correlate horizon dynamics in the strong field regime with waveforms at infinity, paving the way to gravitational tomography {\cite{aa-banff,aank}. For example, in the late time, quasi-normal mode regime, the time evolution of mass multipole moments of the horizon appears to be  strongly correlated with the flux of supermomentum across $\scrip$. This intertwining of observables associated with horizons with those associated with $\scrip$ has the potential for providing fresh insights. In the analysis of isolated gravitating systems in presence of a positive cosmological constant one can use certain cosmological horizons as `local $\scri^\pm$' \cite{Ashtekar_2017,Ashtekar_2019}. The present unified treatment suggests concrete avenues to develop a framework to analyze gravitational waves emitted by these systems and registered at these `local $\scri^\pm$'. In  quantum gravity, this framework is likely to be useful to sharpen the analysis of the black hole evaporation process. It takes some $10^{67}$ years for a solar mass black hole to shrink to lunar mass.  During this very long phase of the evaporation process, the time-evolution of the dynamical horizon should be well-approximated by a perturbed WIH \cite{Ashtekar_2020}. For this regime, the unified framework developed in this paper opens avenues to correlate dynamics of quantum observables defined at the perturbed horizon with those associated with the quantum radiation at $\scrip$. Availability of these two algebras of time-dependent Heisenberg observables will likely deepen our understanding of entanglement (at least) in this semi-classical phase \cite{aa-loops24}. In particular, it would help clarify whether anything dramatic happens to entanglement in the physical space-time at the Page time.

\section*{Acknowledgments}

We thank Neev Khera, Maciej Kolanowski, Jerzy Lewandowski and Antoine Rignon-Bret for discussions. This work was supported in part by the Eberly and Atherton research funds of Penn State and the Distinguished Visiting Research Chair program of the Perimeter Institute. 
We thank the referee for a detailed report that led to improvements in the presentation.

\begin{appendix}
\section{Comparison}
\label{a1}

In this Appendix we compare our method to compute Hamiltonian fluxes
and charges at $\scrip$ to some of the other approaches in the
literature, in particular those based on the Wald-Zoupas (WZ) strategy
\cite{wz,Grant:2021sxk,Odak:2022ndm,Chandrasekaran:2018aop,Odak:2023pga,Chandrasekaran:2023vzb}.
(See also
\cite{Barnich:2011mi,Flanagan:2015pxa,Compere:2018ylh}).
The main differences can be summarized as follow. The
WZ prescription makes
reference not only to the symplectic 2-form, but also to a choice of
symplectic potential for it, and its associated Noether charges.
Consequently, in contrast to the strategy discussed in section
\ref{s3} for $\scrip$, the WZ procedure requires extensions of the BMS
vector fields on $\scrip$ to the space-time interior. On the other
hand, while the WZ procedure can be potentially applied systematically to general covariant theories, it is not clear whether this is possible within our framework. 
For, the simplifications in this framework can be traced back to the fact that one can 
isolate the radiative degrees of freedom at $\scrip$ in general relativity, and this may not be possible for more general gravitational theories.

Given a Lagrangian density 4-form $L$ in space-time, its variation $\d
L\eqons d\big(\Th (\delta)\big)$  provides us with $\Th$ --a 1-form in
the field space and a 3-form in space-time. $\Th$ serves as a
potential for the symplectic current $\J$ --a 2-form in field space
and a 3-form in space-time. Here $\delta$ denotes a vector field on the field 
space as in the main text, and  $\eqons$ stands for
``equals on-shell'', i.e.  when the field equations are satisfied.
We know from the seminal
paper of Emmy Noether that, given a space-time vector field $\xi^a$, and a covariant Lagrangian $L$,
there is a current $j_\xi$ 
that is exact on-shell. In the contemporary language, the 3-form $j_\xi$ is
given by:
\be\label{Noether}
j_\xi:=\Th(\d_\xi) -i_\xi L \eqons d \Q^{(\xi)}
\ee
(see, e.g. \cite{Iyer:1994ys}).
For the Einstein-Hilbert Lagrangian $L$, the standard
choice for $\Th$ is
\be
\Th_{bcd} = \f1{3!} \Th^a \eps_{abcd}, \quad {\rm where} \quad
\Th^a(\d) = \f1{8\pi G}\, g^{a[c}g^{b]d}\, \na_b\d g_{cd}. \label{thg}
\ee
With this choice, $\Q^{(\xi)}=\Q_{\sscr K}^{(\xi)}$ is the Komar 2-form
\be \label{komar}
\Q_{\sscr K}^{(\xi)}{}_{cd} = -\frac 1{16\pi G} \epsilon_{abcd}\nabla^a\xi^b\, .
\ee

Let us now focus on $\scrip$ and restrict $\xi^a$ to be
the BMS vector fields (which are tangential to
$\scrip$). The pull-back to $\scrip$ of \eqref{Noether} gives
\be\label{FB1}
\pbi{d \Q}_{\sscr K}^{(\xi)}\, \eqons \,\pbi{\Th}(\d_\xi).
\ee
If one interprets the 2-forms $\Q^{(\xi)}_{\sscr K}$ as charge aspects, then $\pbi{\Th}(\d_\xi)$ can be interpreted as the flux 3-form:  Eq.~(\ref{FB1}) provides flux-balance laws relating the difference
between charges evaluated on two different 2-sphere cuts of the
hypersurface and the integral of $\Th(\d_\xi)$ over the 3-d region
bounded by them. The problem with this construction is that $\pb{\Th}(\delta_\xi)$ does
not vanish at `non-radiative' solutions $\mathring{g}_{ab}$ of
Einstein's equations for all $\xi$'s (namely solutions with vanishing
news). Therefore, one cannot interpret $\pbi{\Th}(\delta_\xi)$ as the
\emph{physical} flux associated with a BMS vector field $\xi^a$, whence
$\Q^{(\xi)}_{\sscr K}$ cannot be interpreted as the physical charge aspect
either. However, there is an inherent ambiguity in the choice of symplectic potential $\Th$,
and the WZ strategy exploits this freedom to find another one whose charges have better properties.
The key idea is to seek another
symplectic potential $\bar\Th$, a 3-form defined intrinsically at $\scrip$
\be \label{Thetabar}  \b\Th (\delta)\,\, \hat= \,\,\pb{\Th} (\delta) + \delta b, \ee
where $b$ is a 3-form intrinsic to $\scrip$ and chosen so that
$\b\Th$ satisfies locality, analyticity, and covariance, and vanishes
at  non-radiative solutions. The last condition is sometimes referred to as
`stationarity condition', but intended in a looser sense than requiring the 
existence of a time-translation Killing vector.  The importance of these requirements 
can be understood by looking at the 
Hamiltonian one-form associated with a diffeomorphism, given by
\footnote{The term $\Q_{\sscr K}^{(\d\xi)}$ was 
absent in \cite{Iyer:1994ys,wz}, where the identity first appeared.
It was included indirectly in
\cite{Gao:2003ys} and explicitly in
\cite{Barnich:2001jy,Barnich:2011mi}.
The latter references
use a symplectic 2-form that differs by a corner term,
but the difference vanishes in the limit to $\scri$.} 
\be\label{Ixiom1}
\J(\d_\xi,\d)= \d {\Th}(\d_\xi) - \d_\xi{\Th}(\d)
-\Th([\d,\d_\xi])\eqons \rmd \left( \d \Q_{\sscr K}^{(\xi)} -\Q_{\sscr K}^{(\d\xi)} - i_\xi\Th(\d)\right).
\ee
The symplectic current $\J(\d_\xi,\d)$ is independent of the choice of symplectic potential. If we pull it back to $\scrip$ we can replace $\Th$ with $\bar\Th$, 
and field-independence of the BMS vector fields at $\scrip$ guarantees $\bar\Th([\d,\d_\xi])=0$.
Then covariance of $\bar\Th$ guarantees that  $\pbi{\J}(\d_\xi,\d)+di_\xi\bar\Th(\d) = \d \bar\Th(\d_\xi)$ is exact in both spacetime and field-space.
It follows that we can define
\be\label{FB2}
d\Q_{\sscr WZ}^{(\xi)} :\eqons \bar\Th(\d_\xi)
\ee
up to a field-space constant that can be fixed looking at a reference solution.
The flux-balance law is obtained integrating \eqref{FB2} over a region of $\scrip$. The stationarity condition guarantees that the charges are conserved on non-radiative solutions. The covariance requirement is also
crucial to ensure that both the fluxes and the charges reproduce the
symmetry algebra \cite{Rignon-Bret:2024gcx}, removing for instance
field-dependent 2-cocyles as the one found in \cite{Barnich:2011mi}.

In general, there is no guarantee that the required $\b\Theta$
exists in all situations of physical interest, and even if it does, that it is unique.
However at $\scrip$ of asymptotically flat
space-times, $\b\Theta$ has been shown to exist and is uniquely
determined by the conditions above \cite{wz}. Furthermore its
action $\b{\Th}(\delta_\xi)$ is precisely the flux 3-form obtained in
section \ref{s3.2} using Hamiltonian considerations on radiative phase
spaces (see Eq.~(\ref{flux1})):
\be \label{flux}
[\b{\Th} (\delta_\xi)]_{abc}\,\, =\,\, \F^{(\xi)}_{abc}. \ee

Note that \eqref{flux} does not hold for any other symplectic potential. Thus, 
there is unforeseen synergy between our approach and that of WZ.
The origin of this synergy is the fact that $di_\xi\bar\Th$ vanishes precisely on the dense subspace of $\psRh$ defined in section~\ref{s3.2}, and this vanishing occurs \emph{only} if one uses the preferred WZ potential. As a consequence,
\be\label{Ixiomscri}
\bfomegaRh\!\mid_{{}_{\{\hD\}}} (\delta_\xi, \delta) \,=\, 
\d F_\xi[{\cal {\hat R}}] 
\ee
on that dense subspace, hence we can identify the flux as the Hamiltonian generator of the symmetry $\d_\xi$.
We see that the topological argument we used to construct Hamiltonian
fluxes in section~\ref{s3.2} serves two purposes: it
naturally selects the preferred WZ flux, \emph{and} enables us to define a
Hamiltonian on the full radiative phase space $\psRh$ as the continuous
extension from the dense subspace. This provides a precise sense in which the flux $F_\xi [\Rh]$ 
is the generator of the the BMS symmetry $\xi^a$ on the radiative phase space.

It follows from \eqref{flux} that one could just proceed integrating the fluxes as done in the main text to obtain $\Q^{(\xi)}$ of \eqref{Q3}, and identify $\Q_{\sscr WZ}^{(\xi)}\equiv \Q^{(\xi)}$. The WZ paper offered an alternative procedure to determine the charges, based on the introduction of an hyperbolic space-like manifold $\Sigma$ intersecting $\scrip$ at some cross-section $S$, and without internal boundaries. If we pull-back \eqref{Ixiom1} to $\Sigma$ and use \eqref{Thetabar}, we find
\be\label{Ixiom}
\bfomegaSigma(\d_\xi,\d) \eqons \oint_S\left[ \d(\Q_{\sscr K}^{(\xi)}+i_\xi b)-\Q_{\sscr K}^{(\d\xi)}- i_\xi\b\Th(\d)\right].
\ee
Let us suppose that
\be\label{Qdxi}
\Q_{\sscr K}^{(\d\xi)}[S]=\d s_\xi,
\ee
for some $s_\xi$ to be determined.
It then follows from $d\J\eqons 0$ that we can identify
\be\label{QWZ}
\Q_{\sscr WZ}^{(\xi)} = \Q_{\sscr K}^{(\xi)}+i_\xi b- s_\xi,
\ee
up to a field-space constant fixed by the reference solution. Notice that \eqref{QWZ} fixes the ambiguity of adding closed 2-forms to $\Q_{\sscr WZ}^{(\xi)}$, something that one has to do independently in the integrating the fluxes procedure. The formula \eqref{QWZ} has the nice feature of determining $\Q_{\sscr WZ}^{(\xi)}$ in terms of the Komar 2-form and $b$, 
which is arguably simpler than the procedure of integrating the fluxes. However 
the detailed implementation of this strategy requires some care.
If the left-hand side of \eqref{QWZ} is to be identified with the charges obtained integrating the fluxes as in the main text, it has to be independent of the extension of the symmetry vector fields in the bulk. On the right-hand side, we have $i_\xi b$ which is manifestly extension-independent. On the other hand, the integral on cross-sections of $\scrip$
of the Komar 2-form,
\be
 {\Q}_{\sscr K}^{(\xi)}{}_{\pb{cd}} = -\frac 1{16\pi G} \hat
\epsilon_{ab\pb{cd}}\left(\Om^{-2}\hat \nabla^a\hat \xi^b
-2\Om^{-3}\hat n^a\hat \xi^b \right)\, ,
\ee
depends on the $O(\Om^2)$ and $O(\Om^3)$ extension of $\xi^a$ to the bulk, as was already observed in \cite{Geroch:1981ut}. 
Now, one can prove that $\d \Q_{\sscr K}^{(\xi)} - \Q_{\sscr K}^{(\d\xi)}$
depends only on the first-order extension, which is canonical and
field-independent thanks to the universal structure and its embedding in the covariant phase space.
\footnote{A proof when $\d\xi=0$, i.e., when the extension of $\xi^a$ to space-time 
interior is field independent, is given in
\cite{Grant:2021sxk}, Lemma 5.2. It is straightforward to
generalize to $\d\xi\neq 0$ if $\d \Q^{(\xi)}$ is replaced by $\d
\Q^{(\xi)}-\Q^{(\d\xi)}$. Alternatively, with the caveat that $\xi^a$ is always
a c-number for $\d$, as proposed in \cite{Gao:2003ys}.}
Therefore the term $\Q_{\sscr K}^{(\d\xi)}$ compensates the extension dependence of the Komar 2-form. 
We can thus obtain the charges $\Q_{\sscr WZ}^{(\xi)}$ using \eqref{QWZ} provided the assumption \eqref{Qdxi} holds,  a property which is not guaranteed a priori.
If it does, the resulting charges are independent of the symmetry vector field extension, in agreement with the fact that
they match the one derived in the main body of the paper where no
extension was required to begin with.

As an example of implementation of this procedure, let us fix the bulk extension of $\xi$ using the 
Tamborino-Winicour condition \cite{PhysRev.150.1039}. This extension preserves Bondi coordinates in the bulk, 
and it is the most common choice in the literature (see, e.g.,
\cite{Barnich:2011mi,He:2014laa,Campiglia:2015yka,Flanagan:2015pxa,Compere:2018ylh,Compere:2019gft,Campiglia:2020qvc,Elhashash:2021iev,Freidel:2021yqe,Chandrasekaran:2021vyu,Donnay:2022hkf,Geiller:2022vto,Riva:2023xxm}
and references therein). In this case $\d\xi=O(\Om^2)$, and an explicit calculation shows that \eqref{Qdxi} holds with 
\cite{Odak:2022ndm}
\be\label{sxi}
s_\xi:= -\f1{32\pi G}\,\,(\xi^c\hat\ell_c)\,\, ^2\!\hat D_a {}^2\!\hat
D_b\,\hat\s^{ab}\,\, \eps_S\,
\ee
(up to a closed 2-form irrelevant for the charges), 
where $\hat\ell^c$ and its shear $\h\sigma^{ab}$ are adapted to the cross-section.
One can then evaluate the integral of \eqref{QWZ} and verify that it reproduces exactly the BMS charges of the main text. 
Including the contribution of \eqref{sxi} is crucial to obtain this result. 

The bottom line is that once the physical flux is identified, non-trivial work is still required to obtain the charges. One can use the integration procedure of the main text that requires a careful use of Einstein's equations and Bianchi identities, or one can `bootstrap' the Komar expression following the WZ procedure, in which case one
must deal with the subtlety of its extension dependence by including the $\Q_{\sscr K}^{(\d\xi)}$.

\bigskip

\emph{Remarks:} 

1. Looking at the right-hand side of \eqref{Ixiom}, we learn that the
surface charges for vector fields tangent to the corner $S$ are
Hamiltonian generators, whereas those for the remaining vector fields
tangent to $\scrip$ but not to $S$ can only be interpreted as
`generalized Hamiltonians' in the weaker sense that they generate
symmetry transformations at non-radiative solutions where the
`obstruction' $i_\xi\b\Th(\d)$ vanishes. 
The situation is therefore that charges generate all symmetries on the phase space associated to $\Sigma$ only on non-radiative cuts, and the fluxes only 
on dense subspaces of the radiative phase space $\psRh$, which however can be continuously extended to full $\Rh$. 
These caveats are due to the fact that $\scrip$ is a `leaky boundary'. 
In both case the covariant phase space needs to be equipped with information in addition to 
the symplectic 2-form: a prescription for a preferred
symplectic potential, or a topology. \\

2. It follows from  Eq.~\eqref{sxi} that the  contribution $s_\xi$ to the charges
has the structure of a `soft term'. 
Its contribution is non-trivial in that it ensures that the
super-translation charges vanish in Minkowski space-time and that the
boost charges have vanishing flux in the absence of radiation in
divergent-free frames that are not round spheres. \\

3. It is natural to ask if \eqref{FB2} can be understood in terms of Noether's theorem, in other words, if $\Q_{\sscr WZ}^{(\xi)}$ can be derived as an improved Noether charge. The answer is in the affirmative, but the extension dependence of the Komar integral affects also this derivation, making it non-trivial \cite{Odak:2022ndm} (see also \cite{Chandrasekaran:2021vyu,Rignon-Bret:2024gcx}). The extension dependence of the Komar integral makes the pull-back at $\scri$ of $\Th$ anomalous, in the sense that $\d_\xi\pbi{\Th}\neq \pounds_\xi\pbi{\Th}$, and as a consequence $b$ is also anomalous, meaning that $\d_\xi b \neq d i_\xi b$. It is however possible to identify a corner improvement $c$ such that $b':=b+dc$ and $b'$ transforms covariantly under BMS symmetries. Then $\Q_{\sscr WZ}^{(\xi)}$ is the improved Noether charge determined by $\bar\Theta$ and $b'$.\\

4.
There are situations in which there is no symplectic potential $\b\Th$
on $\ps$ satisfying all the requirements of the WZ procedure.
An example is provided by 
asymptotically de Sitter space-times mentioned earlier. In this case,
 one can also carry out
a conformal completion to obtain $\scrip$ which is now space-like. At
$\scrip$, there is a unique $\b{\Th}$ that is selected by the
requirements of locality, analyticity and covariance but it does not
vanish at non-radiative backgrounds;  $\b{\Th}$ does not vanish
already at the Schwartzschild-deSitter space-time \cite{kl2021}.\\

5. There are other instances in which the stationarity condition
required in the WZ procedure is not satisfied by any symplectic
potential for the initial symplectic 2-form. This happens for
conservative boundary conditions at time-like boundaries with
non-orthogonal corners \cite{Harlow:2019yfa,Odak:2021axr}, and for
weaker fall-off conditions at $\scri$ associated with larger symmetry
groups \cite{Compere:2018ylh,Compere:2020lrt,Campiglia:2020qvc,Freidel:2021yqe,Rignon-Bret:2024wlu,Rignon-Bret:2024gcx}.
In these examples it is possible to satisfy the stationarity condition
by allowing in the selection process a larger class of symplectic
potentials, in which a corner term $d\vth$ is added. This changes the
symplectic 2-form, but in a way compatible with the field equations
and part of the covariant phase space ambiguities
\cite{Jacobson:1993vj,Compere:2008us}. Of this type are for instance
the differences between the Einstein-Hilbert symplectic 2-form and the
ADM one \cite{Burnett:1990tww} or the tetrad one
\cite{DePaoli:2018erh}.\\

6. Another situation to consider is when the stationarity condition is
too restrictive for some physical applications. For instance the WZ
prescription applied to arbitrary null hypersurfaces in \cite{cfp}
treats as stationary only shear and expansion-free null surfaces
\cite{Chandrasekaran:2018aop}. As pointed out in
\cite{Ashtekar:2021kqj}, it excludes situations that are manifestly
non-radiative, such as a null cone in Minkowski. This shortcoming can
be dealt with by weakening the stationarity requirement to be the
vanishing of the Noether flux $\bar\Th(\d_\xi)$, as
opposed to the vanishing of the full symplectic flux, and this leads
to select a different and again unique symplectic potential whose
charges are conserved on both non-expanding horizons and null light
cones \cite{Rignon-Bret:2023fjq,Odak:2023pga}.\\

For other approaches to charges at $\scrip$ see also
\cite{Campiglia:2015yka,Barnich:2016lyg,Compere:2017wrj,Barnich:2019vzx,Prabhu:2019fsp,Wieland:2019hkz,Wieland:2020gno,Campiglia:2020qvc,Wieland:2021eth,Freidel:2021yqe,Freidel:2021ytz,Chandrasekaran:2021vyu,Donnay:2022hkf,Geiller:2022vto}.

\end{appendix}

\providecommand{\href}[2]{#2}\begingroup\raggedright\endgroup


\end{document}